\PassOptionsToPackage{hyphens}{url}
\documentclass[a4paper,fleqn]{cas-dc}

\usepackage[authoryear]{natbib}
\usepackage{xurl}
\usepackage{hyperref}
\usepackage{placeins} 
\usepackage{tikz}
\usetikzlibrary{arrows.meta,fit,positioning}

\AtBeginEnvironment{thebibliography}{%
  \sloppy\emergencystretch=2em
  \Urlmuskip=0mu\relax
  \clubpenalty=0\widowpenalty=0\brokenpenalty=0\interlinepenalty=0\relax
  \def\URLprefix{\ifhmode\unskip\space\fi URL:\space}%
  \def\DOIprefix{\ifhmode\unskip\space\fi doi:\space}%
}

\ExplSyntaxOn
\RenewDocumentCommand \printorcid { }
  {
    \seq_if_empty:NF \g_stm_orcid_seq
      {
        \group_begin:
          \tex_let:D \thefootnote \relax \footnotetext
          {
            \raggedright
            \textsc{orcid}(s):\c_space_token
            \seq_use:Nn \g_stm_orcid_seq { ;~ }
          }
        \group_end:
      }
  }
\ExplSyntaxOff

\newcommand{\best}[1]{\begingroup\bfseries\boldmath #1\endgroup}
\newcommand{\schemabox}[2]{\textbf{#1}\\[-0.15em]\rule{0.88\linewidth}{0.25pt}\\[-0.05em]#2}
\newcolumntype{V}{!{\vrule width 0.5pt\hspace{2.2pt}\vrule width 0.5pt}}
\ExplSyntaxOn
\NewDocumentCommand{\cmcell}{m}
  {
    \group_begin:
      \int_set:Nn \l_tmpa_int { \fp_eval:n { round(max(0,min(100,#1)),0) } }
      \int_set:Nn \l_tmpb_int { \int_eval:n { 100 - \l_tmpa_int } }
      \cellcolor{black!\int_use:N \l_tmpa_int !white}%
      \textcolor{black!\int_use:N \l_tmpb_int !white}{\rule{0pt}{2.1ex}#1\%}%
    \group_end:
  }
\ExplSyntaxOff

\def\tsc#1{\csdef{#1}{\textsc{\lowercase{#1}}\xspace}}
\tsc{WGM}
\tsc{QE}

\begin{document}
\let\WriteBookmarks\relax
\renewcommand{\floatpagefraction}{0.9}
\renewcommand{\dblfloatpagefraction}{0.9}
\renewcommand{\textfraction}{0.01}
\renewcommand{\topfraction}{0.95}
\renewcommand{\dbltopfraction}{0.95}
\makeatletter
\setlength{\@dblfptop}{0pt}
\setlength{\@dblfpsep}{12pt}
\setlength{\@dblfpbot}{0pt plus 1fil}
\makeatother

\shorttitle{STRADAViT}    

\shortauthors{A. DeMarco et~al.}

\title [mode = title]{STRADAViT: Self-Supervised Domain Adaptation of Vision Transformer Backbones for Radio Astronomy}

\nonumnote{Authors are listed in order of contribution; the final author provided advisory support during project planning and funding application.}

%

\author[1]{Andrea {DeMarco}}
\cormark[1]
\ead{andrea.demarco@um.edu.mt}
\ead[url]{https://www.um.edu.mt/profile/andreademarco}
\credit{Conceptualization, Methodology, Software, Data curation, Formal analysis, Investigation, Visualization, Writing -- original draft, Writing -- review \& editing, Project administration, Funding acquisition}

\cortext[1]{Corresponding author}

\affiliation[1]{organization={Institute of Space Sciences and Astronomy},
            addressline={University of Malta}, 
            city={Msida},
            postcode={MSD2080}, 
            country={Malta}}

\affiliation[2]{organization={Istituto Nazionale di Astrofisica (INAF)},
            addressline={Osservatorio Astrofisico di Catania}, 
            city={Catania},
            country={Italy}}

\affiliation[3]{organization={School of Physics and Astronomy, Royal Observatory},
            addressline={University of Edinburgh}, 
            city={Edinburgh},
            country={United Kingdom}}

\author[1]{Ian {Fenech Conti}}
\credit{Data curation, Validation, Writing -- review \& editing}

\author[1]{Hayley {Camilleri}}
\credit{Data curation, Validation, Writing -- review \& editing}

\author[3]{Ardiana {Bushi}}
\credit{Data curation}

\author[2]{Simone {Riggi}}
\credit{Validation, Writing -- review \& editing}



\begin{abstract}
Next-generation radio astronomy surveys are delivering millions of resolved sources, yet scalable morphology analysis remains difficult across heterogeneous telescopes and imaging pipelines.
We present STRADAViT, a self-supervised continued-pretraining framework for learning transferable radio-astronomy encoders
from Vision Transformer (ViT) backbones. It combines mixed-survey data curation, radio astronomy-aware training-view
generation, and a ViT-MAE-initialized encoder family with optional register tokens. It supports reconstruction-only, contrastive-only, and
two-stage branches. Our pretraining dataset comprises
$512\times512$ radio astronomy cutouts drawn from four complementary sources (MeerKAT, ASKAP, LOFAR/LoTSS, and SKA SDC1 simulated data). We evaluate transfer with linear probing (LP) and fine-tuning (FT) on three
morphology benchmarks spanning binary and multi-class settings (MiraBest, LoTSS DR2, and Radio Galaxy Zoo). An exploratory
three-fold ablation grid guides selection of a register-based two-stage checkpoint using a fixed cross-dataset criterion.
Across subsequent 15-seed paired downstream evaluations on fixed partitions, this checkpoint improves linear-probe Macro-F1
over its ViT-MAE initialization on all three benchmarks and improves fine-tuning on MiraBest and RGZ DR1, while LoTSS DR2
fine-tuning declines; all six differences remain statistically supported after Holm correction. A parallel DINOv2 experiment
yields mixed adaptation effects: the procedure transfers, but the benefit is not uniform. STRADAViT thus
improves frozen ViT representations while retaining clear dataset-dependent limitations and remaining below task-specialized
methods on standard MiraBest classification.
\end{abstract}




\maketitle

\section{Introduction}\label{intro}
Radio astronomy is entering an era of high-volume imaging surveys, with instruments such as MeerKAT, 
ASKAP and LOFAR producing increasingly deep and wide observations, and future facilities such as the 
SKA expected to extend both scale and sensitivity. These surveys contain diverse morphologies and 
imaging artifacts that can masquerade as morphology. The scientific return is often constrained by the availability of reliable, scalable image-based analysis tools that 
generalize across instruments and data products, rather than by raw data volume.

In radio astronomy, supervised deep learning faces two recurring constraints. First, high-quality 
labels for morphology are expensive and typically derived from expert visual inspection or citizen 
science. Second, models trained on one survey or imaging pipeline can degrade when applied to others 
due to differences in angular resolution, $uv$-coverage, deconvolution, noise statistics, dynamic 
range, and calibration/imaging conventions. This motivates approaches that exploit abundant unlabeled 
radio astronomy data while explicitly targeting cross-telescope robustness.

Self-supervised pretraining offers a way to learn transferable Vision Transformer (ViT) representations without
morphology labels. For radio astronomy, the challenge is adapting these methods to single-channel data,
survey-dependent intensity distributions, and heterogeneous instrumental and imaging systematics.

This paper presents STRADAViT, a self-supervised continued-pretraining framework for radio astronomy ViT backbones. The study
deliberately restricts its architectural scope to ViT and ViT-derived backbones in order to isolate the effects of mixed-survey
continued pretraining, radio astronomy-aware view generation, and objective design across heterogeneous morphology benchmarks.
Our central question is whether region-of-interest (ROI)-aware view generation and staged reconstruction-to-contrastive continued pretraining can produce
ViT representations that transfer more effectively across heterogeneous radio morphology benchmarks.

Our primary contributions are:
\begin{itemize}
\item a mixed-survey $512\times512$ radio astronomy cutout dataset spanning four complementary sources;
\item radio astronomy-aware training-view generation that anchors self-supervised views to informative source regions instead of relying on naive random crops;
\item a controlled continued-pretraining framework initialized from ViT-MAE, with optional register tokens and controlled ablations of reconstruction losses, hard-negative-aware contrastive losses, and reconstruction-only, contrastive-only, and two-stage training branches;
\item a Soft-HCL variant that mixes uniform and similarity-weighted negative distributions to limit concentration on the hardest negatives;
\item a targeted DINOv2 initialization ablation under fixed HCL training, assessing whether continued pretraining extends beyond ViT-MAE under the same downstream protocol;
\item a reproducible transfer protocol (linear probing and fine-tuning) across three public morphology benchmarks, combining
exploratory three-fold ablations with 15-seed paired comparisons of fixed STRADAViT and DINOv2 checkpoints against their
corresponding initializations.
\end{itemize}

For frozen encoders, the selected release checkpoint improves Macro-F1 over ViT-MAE on all three datasets, with statistically supported
paired differences across downstream seeds on fixed partitions. Fine-tuning improves on MiraBest and RGZ DR1 but declines
on LoTSS DR2. Improved representation quality, then, does not amount to uniform downstream superiority.

Figure~\ref{fig:branch_overview} summarizes the overall workflow studied in this paper, from mixed-survey continued
pretraining through the three training branches to downstream transfer evaluation.

\begin{figure*}[tb]
\centering
\resizebox{\textwidth}{!}{%
\begin{tikzpicture}[
  font=\sffamily\scriptsize,
  node distance=7mm and 8mm,
  block/.style={
    draw=black!60,
    rounded corners=2pt,
    line width=0.35pt,
    fill=black!3,
    align=center,
    inner xsep=5pt,
    inner ysep=5pt,
    minimum height=14mm
  },
  source/.style={block, fill=black!2, text width=0.17\textwidth},
  view/.style={block, fill=green!5, text width=0.17\textwidth},
  branch/.style={block, fill=blue!4, text width=0.18\textwidth},
  output/.style={block, fill=orange!6, text width=0.15\textwidth},
  eval/.style={block, fill=purple!5, text width=0.17\textwidth},
  group/.style={
    draw=black!35,
    rounded corners=3pt,
    dashed,
    inner xsep=5pt,
    inner ysep=7pt
  },
  arrow/.style={-{Latex[length=2.0mm]}, draw=black!70, line width=0.45pt}
]
\node[source] (data) {\schemabox{Pretraining data}{Sources: MeerKAT, ASKAP\\LoTSS, SKA SDC1\\Input: $512\times512$ radio cutouts}};
\node[view, right=of data] (views) {\schemabox{View pipeline}{Standardize: ZScale\\Generate: augmented views\\Sampling: ROI-anchored\\single or paired views}};

\node[branch, right=14mm of views, yshift=24mm] (recon) {\schemabox{Reconstruction-only}{Task: masked-patch\\reconstruction\\Loss: L2, L2+L1\\L2+BL1}};
\node[branch, right=14mm of views] (contrastive) {\schemabox{Contrastive-only}{Views: two correlated\\augmented views\\Loss: InfoNCE, HCL\\Soft-HCL}};
\node[branch, right=14mm of views, yshift=-24mm] (twostage) {\schemabox{Two-stage}{Reconstruction warm start\\followed by\\contrastive refinement}};
\node[group, fit=(recon)(contrastive)(twostage), label={[font=\sffamily\scriptsize]above:Continued-pretraining branches}] (branches) {};

\node[output, right=14mm of contrastive] (encoders) {\schemabox{Adapted ViT encoders}{ViT-MAE family\\or\\DINOv2 HCL\\initialization ablation}};
\node[eval, right=of encoders] (transfer) {\schemabox{Transfer evaluation}{Protocol: linear probing\\or full fine-tuning\\Benchmarks: MiraBest\\LoTSS DR2, RGZ DR1}};

\draw[arrow] (data) -- (views);
\draw[arrow] (views.east) -- (recon.west);
\draw[arrow] (views.east) -- (contrastive.west);
\draw[arrow] (views.east) -- (twostage.west);
\draw[arrow] (recon.east) -- (encoders.west);
\draw[arrow] (contrastive.east) -- (encoders.west);
\draw[arrow] (twostage.east) -- (encoders.west);
\draw[arrow] (encoders) -- (transfer);
\end{tikzpicture}
}
\caption{High-level overview of STRADAViT. Mixed-survey radio cutouts are standardized and converted into radio astronomy-aware training views, then used in reconstruction-only, contrastive-only, or two-stage continued-pretraining branches. The adapted encoders are evaluated under the same downstream linear-probing and fine-tuning protocol across the three morphology benchmarks.}\label{fig:branch_overview}
\end{figure*}

We describe the datasets, training objectives, and transfer protocol before comparing the training branches and
their limitations. Appendix~\ref{app:mirabest_standard_split} provides the standard-split MiraBest comparison with prior work.

\section{Literature Survey}\label{litreview}
Convolutional networks remain strong imaging baselines \citep{He2016ResNet}, while ViTs extend the Transformer
architecture to large-scale visual representation learning \citep{Vaswani2017Attention,Dosovitskiy2021ViT}.
The experiments are restricted to ViT architectures to isolate the effects of pretraining and view generation.

Contrastive methods learn invariances across augmented views \citep{Chen2020SimCLR,He2020MoCo}; related approaches
alter negative handling or regularize feature statistics \citep{Grill2020BYOL,Bardes2022VICReg}.
MAE instead reconstructs masked image content \citep{He2022MAE}, motivating the reconstruction-to-contrastive
sequence studied here. DINO and DINOv2 demonstrate strong self-supervised ViT transfer
\citep{Caron2021DINO,Oquab2023DINOv2}. Although general-purpose pretraining yields transferable features
\citep{Radford2021CLIP}, radio astronomy requires normalization and augmentations that preserve morphology
across survey-dependent noise, dynamic range, and artifacts.

\subsection{Radio Astronomy Context and Benchmark Labels}
Beyond morphology classification, radio astronomy has a long tradition of source finding and characterization under
survey-specific artifacts and systematics, as illustrated by CAESAR and related practical pipelines
\citep{Riggi2016CAESAR,Riggi2019CaesarPASA}. Recent work extends this toward broad reviews of deep learning for radio
classification \citep{Riggi2024DeepLearningRadio}, self-supervised transfer and benchmark studies
\citep{Riggi2024PASAContrastive,Cecconello2025RadioSSLBenchmark}, and multimodal systems
\citep{Riggi2025RadioLLaVA,Drozdova2025VLMRadio}. Additional SSL studies now cover source classification, FR\,I/FR\,II
transfer, large-scale pretraining on Radio Galaxy Zoo, latent-space radio-galaxy representations, and self-supervised
learning on MeerKAT continuum images
\citep{BaronPerez2025AandASSL,Buatthaisong2025MNRASSelfSupervisedFR,Slijepcevic2024RGZFoundation,Andrianomena2023RGZVAE,Lastufka2024MeerKATSSL}.

Our benchmarks span curated FR\,I/FR\,II labels in MiraBest \citep{PorterScaife2023MiraBest}, extended-source
morphology in LoTSS DR2 \citep{Horton2025LoTSSDR2Morphology}, and citizen-science component/peak labels in
RGZ DR1 \citep{RGZDR1Zenodo14195049}. Their label spaces are not harmonized across surveys
(Section~\ref{classification_data}).

\subsection{Positioning Relative to Recent Radio Astronomy SSL Literature}
Relative to recent radio astronomy SSL and representation studies
\citep{Riggi2024PASAContrastive,Cecconello2025RadioSSLBenchmark,Lastufka2024MeerKATSSL,
BaronPerez2025AandASSL,Buatthaisong2025MNRASSelfSupervisedFR,Slijepcevic2024RGZFoundation,Andrianomena2023RGZVAE}, STRADAViT emphasizes
four points: mixed-survey pretraining, ROI-aware view generation for sparse cutouts, a staged
reconstruction-to-contrastive objective, and a controlled ViT-only architectural scope. The aim is cross-survey reuse across
multiple downstream label spaces rather than optimization for a single benchmark. Task-specialized MiraBest studies provide
a complementary external reference \citep{Hossain2023SemiSupervisedGCNN,Bowles2021Attention,ScaifePorter2021EquivariantCNN}.
Appendix~\ref{app:mirabest_standard_split}
places the MiraBest results in the context of the corresponding prior-work protocols. Direct numerical comparison with
published work is restricted to MiraBest because the LoTSS DR2 and RGZ DR1 studies considered here do not use the same
sample selection, label construction, input definition, and evaluation protocol.

\section{Unsupervised Learning Dataset}\label{ssl_data}
We assemble $512\times512$ continuum-image cutouts from MeerKAT MGCLS DR1 \citep{Knowles2022MGCLS}, ASKAP images accessed via the CSIRO ASKAP
Science Data Archive (CASDA)\footnote{\url{https://casda.csiro.au/casda_data_access/}}, LOFAR LoTSS mosaics \citep{Shimwell2022LoTSS}, and simulations from the first SKA Science Data Challenge (SDC1) DR1
\citep{Bonaldi2020}. These sources expose the model to different angular resolutions, noise properties, artifacts,
and morphologies. Table~\ref{tab:ssl_corpus} gives the final cutout inventory after image preprocessing, filtering,
and the spatial exclusion described in Section~\ref{spatial_exclusion}.

\subsection{Cutout Extraction and Basic Filtering}
All inputs are treated as single-channel images stored in FITS. To obtain a uniform pixel-space resolution, we tile each survey 
image into non-overlapping $512\times512$ cutouts (stride equal to tile size), discarding partial tiles at image edges. 
When the input product is a cube, we iterate over its leading axes and extract cutouts from each 2D plane.

We apply basic filtering to exclude uninformative or corrupted data from the pretraining set. 
Cutouts containing NaN/Inf values or consisting entirely of zeros are discarded. We additionally skip malformed FITS products 
(e.g., absent or empty image arrays, or non-numeric pixel types).

\subsection{Cutout Normalization}
For morphology-focused learning, each cutout is independently stretched with Astropy's \texttt{ZScaleInterval} \citep{Astropy2013}
(IRAF-style ZScale), linearly mapped using the estimated lower/upper limits, and clipped to $[0,1]$.
Non-finite values are set to zero; pathological constant cutouts map to zeros.
This deliberately discards absolute flux calibration. The scaled single channel is replicated to three channels
for compatibility with standard vision backbones.

\begin{table}[pos=t]
\caption{Composition of the self-supervised pretraining dataset (all images are $512\times512$ cutouts). The LoTSS count is after spatial exclusion. ``Ratio'' denotes the fraction of total cutouts contributed by each source.}\label{tab:ssl_corpus}
\centering
\begin{tabular*}{\tblwidth}{@{}@{\extracolsep{\fill}}lrr@{}}
\toprule
Source & Cutouts & Ratio \\
\midrule
MGCLS (MeerKAT), DR1 & 139500 & 0.24 \\
ASKAP (CASDA) & 261390 & 0.44 \\
LoTSS (LOFAR) & 177476 & 0.30 \\
SKA SDC1, DR1 & 12288 & 0.02 \\
\midrule
Total & 590654 & 1.00 \\
\bottomrule
\end{tabular*}
\end{table}

\subsection{Online View Generation}
Training views are generated on-the-fly, exposing each cutout to varying ROI selections, crop scales, and augmentations
(Sections~\ref{phase1} and \ref{phase2}). Additional surveys can enter through the same FITS preprocessing interface.

\subsection{Survey Overlap and Spatial Exclusion}\label{spatial_exclusion}
The direct survey-level overlap risk arises for LoTSS DR2 because LoTSS mosaics also contribute to the self-supervised
pretraining dataset. To avoid leakage in this case, labeled LoTSS DR2 evaluation sources were excluded from the pretraining
manifest at the sky-coordinate level: candidate pretraining cutouts were checked against the labeled source coordinates and
removed when an evaluation-source coordinate fell inside the cutout's celestial WCS footprint, using zero additional pixel
padding. Zero padding defines exclusion by the measured WCS footprint, providing a consistent coordinate-based rule
without imposing a common angular extent across morphologically diverse sources.
MiraBest uses FIRST and NVSS imagery, and RGZ DR1 uses FIRST imagery \citep{Becker1995FIRST}; neither survey is included
among the pretraining image products. The role of evaluation samples in checkpoint selection is described in
Section~\ref{post_selection_eval}.

\section{Evaluation Datasets}\label{classification_data}
We evaluate transfer learning on three public radio astronomy imaging datasets that span different telescopes,
labeling paradigms, and class granularities: MiraBest \citep{PorterScaife2023MiraBest}, the LoTSS DR2 visual-classification
sample of \citet{Horton2025LoTSSDR2Morphology}, and Radio Galaxy Zoo DR1 \citep{RGZDR1Zenodo14195049}. Together, these datasets
probe compact binary morphology, richer multi-class morphology, and component/peak complexity under heterogeneous imaging
products.

Table~\ref{tab:eval_datasets} summarizes the number of retained samples in each evaluation dataset after applying the dataset
selection and filtering criteria described below.
\begin{table}[pos=t]
\caption{Evaluation datasets and retained image counts for the exploratory grid. MiraBest combines FIRST and NVSS image entries; the count is not a count of distinct sources.}\label{tab:eval_datasets}
\centering
\begin{tabular*}{\tblwidth}{@{}@{\extracolsep{\fill}}lr@{}}
\toprule
Dataset & Samples \\
\midrule
Radio Galaxy Zoo DR1 & 98,391 \\
MiraBest & 1,563 \\
LoTSS DR2 & 8,805 \\
\bottomrule
\end{tabular*}
\end{table}

\begin{figure*}[tb]
\centering
\includegraphics[width=\textwidth,height=0.60\textheight,keepaspectratio]{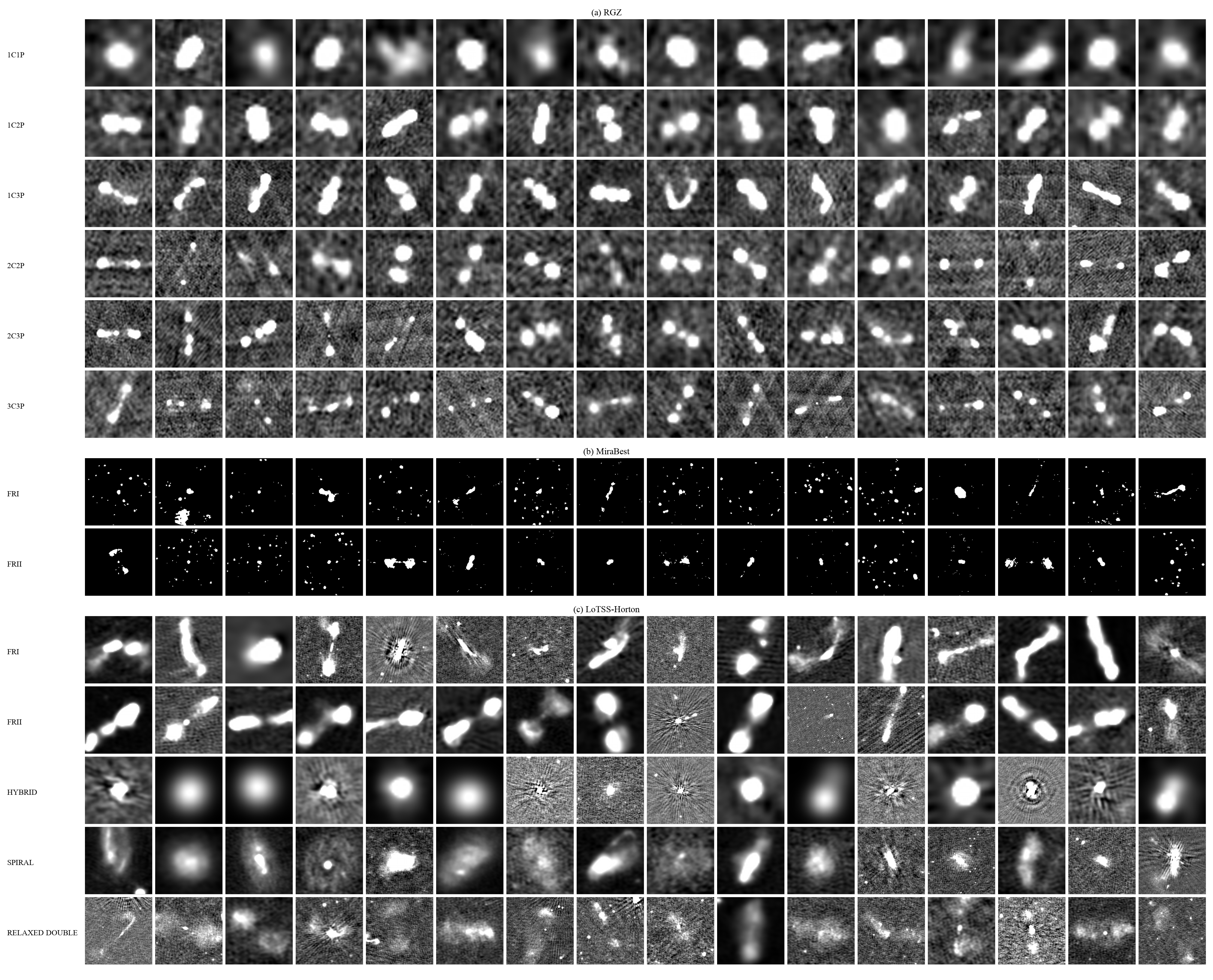}
\caption{Example cutouts from the three evaluation datasets after preprocessing with the same pipeline used for pretraining and evaluation (Section~\ref{ssl_data}). Panels show samples of each class for (a) RGZ DR1, (b) MiraBest, and (c) the LoTSS DR2 visual-classification sample of \citet{Horton2025LoTSSDR2Morphology}.}\label{fig:test_views}
\end{figure*}

Figure~\ref{fig:test_views} shows representative, preprocessed examples from each dataset (panels (a)--(c)), illustrating the
visual diversity of the three evaluation benchmarks under a shared input pipeline.

\subsection{Common Preprocessing}
Evaluation images use the per-image ZScale, non-finite-value handling, and three-channel replication described in
Section~\ref{ssl_data}, followed by bicubic resizing to $224\times224$ for ViT-MAE-family models or the native
$518\times518$ for DINOv2-family models.

\subsection{MiraBest (Binary FR\,I/FR\,II)}
MiraBest provides a curated set of morphologically classified radio galaxies intended for machine learning applications
\citep{PorterScaife2023MiraBest}. The released labels include multiple subtypes and confidence levels. For this work we use
both the F (FIRST) and N (NVSS) batches distributed with MiraBest v2\footnote{\url{https://doi.org/10.5281/zenodo.5588282}}, and cast the benchmark as a binary Fanaroff--Riley task by
mapping all FR\,I subtypes to FR\,I and all FR\,II subtypes to FR\,II, and exclude hybrid sources. Unless stated otherwise,
we further restrict to confident labels and discard uncertain entries to avoid conflating representation quality with label
ambiguity. This retained set is used for the main cross-validated transfer evaluation. Direct comparison with prior MiraBest
accuracy or test-error studies is handled separately in Appendix~\ref{app:mirabest_standard_split}, which uses the fixed
MiraBest Confident standard train/test split. F and N contain observations of the same source catalog at different
pixel scales (1.8 and 15 arcsec per pixel, respectively). The 1,563 entries reported here are retained images, not
1,563 distinct confident galaxies; the recorded class counts are 745 FR\,I and 818 FR\,II. Class balancing for
Figure~\ref{fig:representation_umap} retains 745 per class, hence 1,490 images. The standard Dataset-F comparison instead
contains 833 images. Exploratory folds pool the available training and test batches and are stratified by image rather
than source, allowing FIRST/NVSS counterparts to cross folds. The 104 standard Dataset-F test images were therefore
not reserved from exploratory checkpoint selection.

\subsection{LoTSS DR2 (Multi-Class Initial Labels)}
\citet{Horton2025LoTSSDR2Morphology} present a large visual-classification effort for bright, resolved LoTSS DR2 sources,
including an ``initial'' class taxonomy and additional morphology and environment flags. We download high-resolution LoTSS DR2
cutouts for each source using the public cutout service, centering on the catalog coordinates. Cutout angular size is computed
per source from the reported redshift and physical size when available; otherwise we fall back to a fixed cutout size (5
arcmin) and do not apply additional enlargement (factor $=1$). Specifically, the requested side length is
$\theta_{\rm arcmin}=[L_{\rm kpc}/D_{\rm A,kpc}(z)](180/\pi)60$, using the catalog physical size $L$ and Astropy's
Planck18 angular-diameter distance $D_{\rm A}$ \citep{Planck2020Parameters}.

To keep the benchmark label space well-defined, we use only the initial class taxonomy (FR\,I, FR\,II, Hybrid, Spiral, and
Relaxed double) and restrict to sources with exactly one active initial label. Samples with no initial label (or ambiguous
multiple initial labels) are excluded.

\subsection{Radio Galaxy Zoo DR1 (Multi-Class Component/Peak Morphology)}
Radio Galaxy Zoo DR1 provides citizen-science classifications for complex radio sources, including a consensus level that can
be used to filter for high-confidence labels \citep{RGZDR1Zenodo14195049}. From the FIRST subset, we construct a multi-class
benchmark based on the discrete component/peak summary provided in the catalog. Specifically, we derive a class label from the
number of radio components and peaks and retain the six most common combinations:
1C1P, 1C2P, 1C3P, 2C2P, 2C3P, and 3C3P. We filter the catalog to consensus level $\ge 0.65$ and download cutouts via the
FIRST cutout service, using an adaptive angular size of 1.5 times the catalog largest angular extent (LAE) and
enforcing a minimum cutout side length of 8 pixels. To improve data integrity, cutouts are rejected when the WCS-derived image
center deviates from the target coordinate by more than 1 arcmin.

\section{ViT Backbone Architecture}\label{vit_backbone}
We retain a ViT-MAE base encoder and compare optional projection heads and register tokens.

\subsection{Base Encoder: ViT-MAE}
STRADAViT uses the standard ViT-MAE base encoder (ViT-B/16) and lightweight reconstruction decoder
\citep{He2022MAE,Dosovitskiy2021ViT,Vaswani2017Attention}. The encoder uses $16\times16$ patches, 12 Transformer
blocks, 12 attention heads, and embedding dimension $D=768$; the standard MAE decoder architecture is unchanged.
This provides a common encoder for reconstruction and contrastive branches without introducing architectural variation
between those branches.

Reconstruction uses the decoder to predict masked patches; contrastive training uses the encoder alone.
For MAE-family contrastive and downstream readout, we mean-pool patch tokens, excluding CLS and register tokens,
because the reconstruction objective optimizes patch content rather than a CLS representation.
Concretely, if $h_i\in\mathbb{R}^D$ denotes the last-layer embedding of the $i$th patch token (with $i=1,\ldots,N$), we form
\begin{equation}
z=\frac{1}{N}\sum_{i=1}^{N} h_i,
\end{equation}
and use $z$ as the per-view representation. Contrastive loss acts on the
projection of $z$, not the CLS token. DINOv2 adaptation instead feeds the backbone's final LayerNorm-normalized CLS
token (the Hugging Face \texttt{pooler\_output}) to the projection head. The same CLS representation feeds its downstream
linear classifier; it is not concatenated with mean patch tokens.

\subsection{Projection Head for Contrastive Learning}
Contrastive pretraining uses an additional projection head that maps the pooled encoder embedding to a lower-dimensional space
in which the contrastive objective is applied. We use a two-layer MLP with hidden dimension 2048 and output dimension 128.
For ViT-MAE, the MLP uses LayerNorm after the first linear layer, a ReLU nonlinearity, and Batch Normalization on the output. The resulting
vectors are $\ell_2$-normalized before computing the contrastive loss. We train on 4 GPUs and synchronize the projection-head
BatchNorm statistics across devices (SyncBatchNorm). The DINOv2 projector omits output BatchNorm.

\subsection{Register Tokens}
We also evaluate an optional register-token variant inspired by register-augmented ViT designs \citep{Darcet2024Registers}.
In this setting, we insert $R$ learnable register tokens immediately after the class token and before the patch tokens. These
register tokens participate in all encoder blocks and can act as additional content-bearing slots, while leaving the patch token
grid unchanged. For compatibility with masked reconstruction and for consistent downstream readout, register tokens are removed
from the sequence before the reconstruction decoder and before patch-token pooling in the contrastive objective. We report results for
$R\in\{0,4\}$, where $R=0$ recovers the standard ViT-MAE encoder.

\subsection{Continued-Pretraining Branches}
We consider three continued-pretraining branches. The \emph{reconstruction-only} branch applies the masked reconstruction
objective by itself. The \emph{contrastive-only} branch applies the contrastive objective directly from the chosen initial
checkpoint, without a preceding reconstruction stage. The \emph{two-stage} branch first trains with the
reconstruction objective and then continues from that checkpoint with the contrastive objective. We use ``branch'' for these
pipelines and reserve ``phase~1'' and ``phase~2'' for the two stages of the two-stage branch.

\section{Masked Reconstruction Objective}\label{phase1}

\subsection{View Generation for Masked Reconstruction}
Each standardized $512\times512$ cutout yields one reconstruction view. ROI-aligned cropping limits empty,
background-dominated views while retaining a controlled fraction of wide-field context.

\begin{figure*}[tb]
\centering
\includegraphics[width=\textwidth,height=0.60\textheight,keepaspectratio]{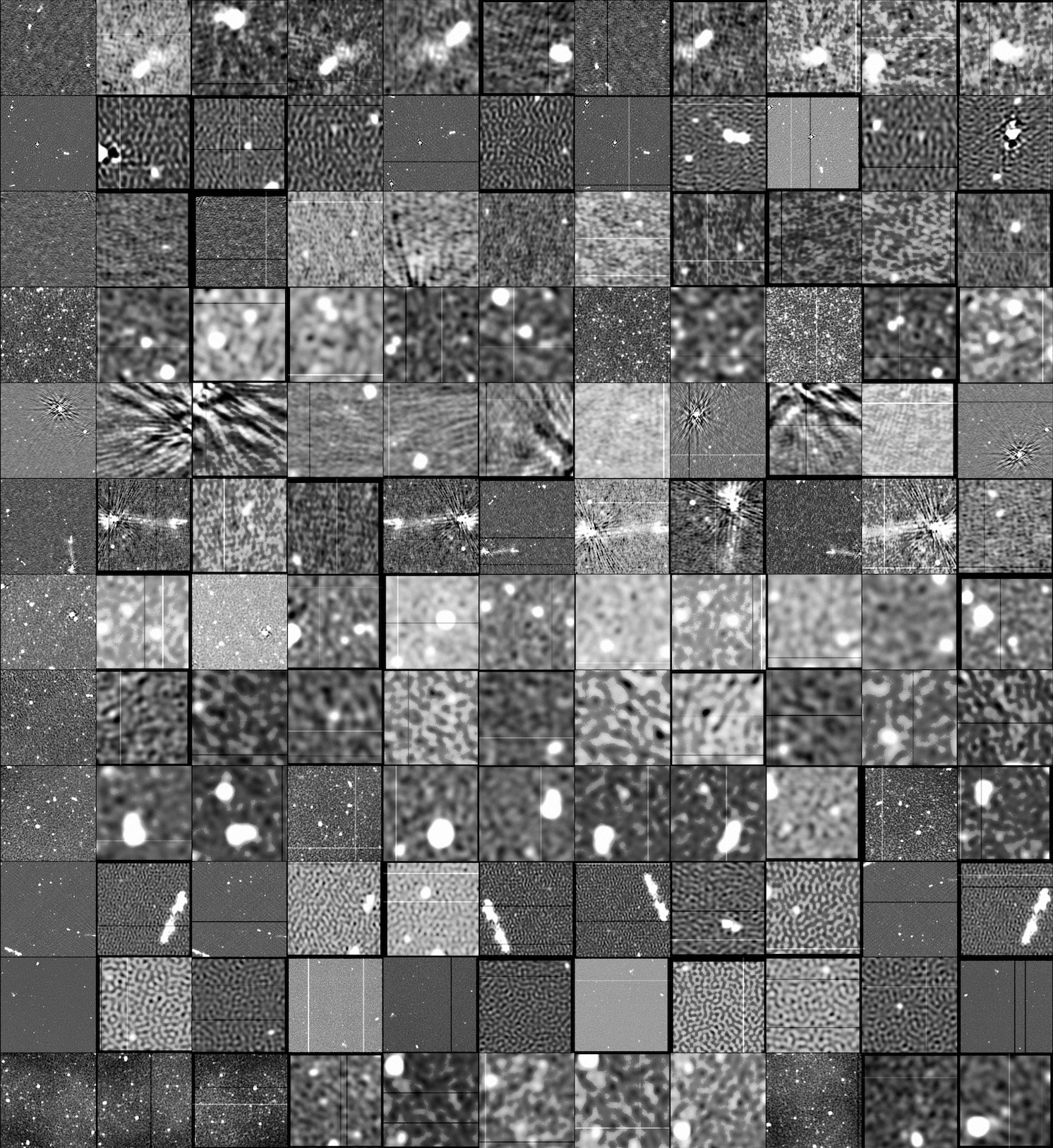}
\caption{Reconstruction-branch masked-reconstruction views. The first column shows the standardized parent cutout (after the per-image
ZScale contrast stretch described in Section~\ref{ssl_data}). Subsequent columns show example ROI-aligned crops produced by our
single-view strategy (Section~\ref{phase1}); with probability $p_{\mathrm{global}}=0.2$ the strategy instead samples a
wide-field crop from the full cutout. The selected view is then transformed by progressively applied, morphology-preserving
augmentations.}\label{fig:mae_views}
\end{figure*}

\begin{figure*}[tb]
\centering
\begin{minipage}[t]{0.49\textwidth}
\centering
\includegraphics[width=\linewidth]{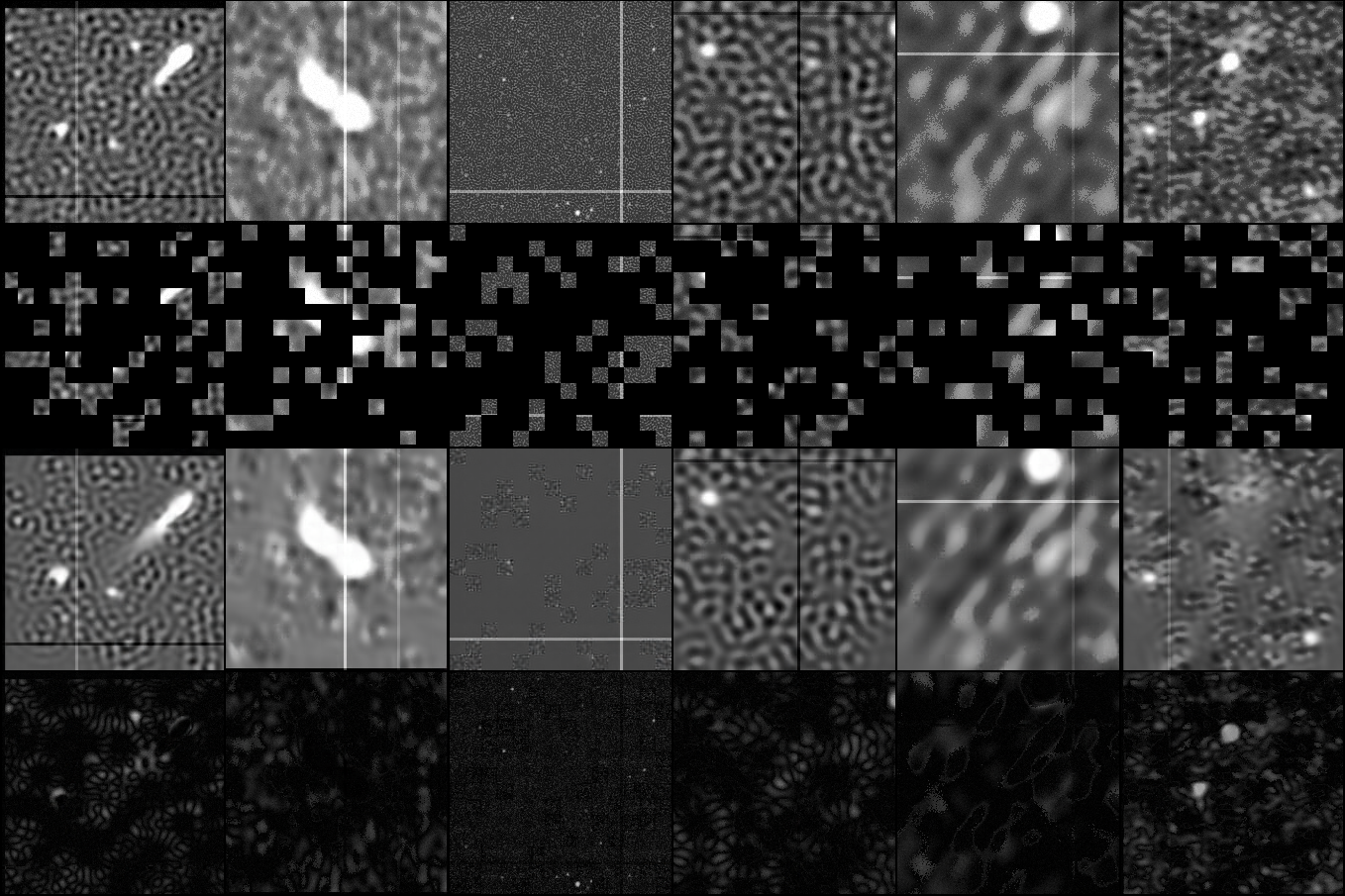}
\end{minipage}\hfill
\begin{minipage}[t]{0.49\textwidth}
\centering
\includegraphics[width=\linewidth]{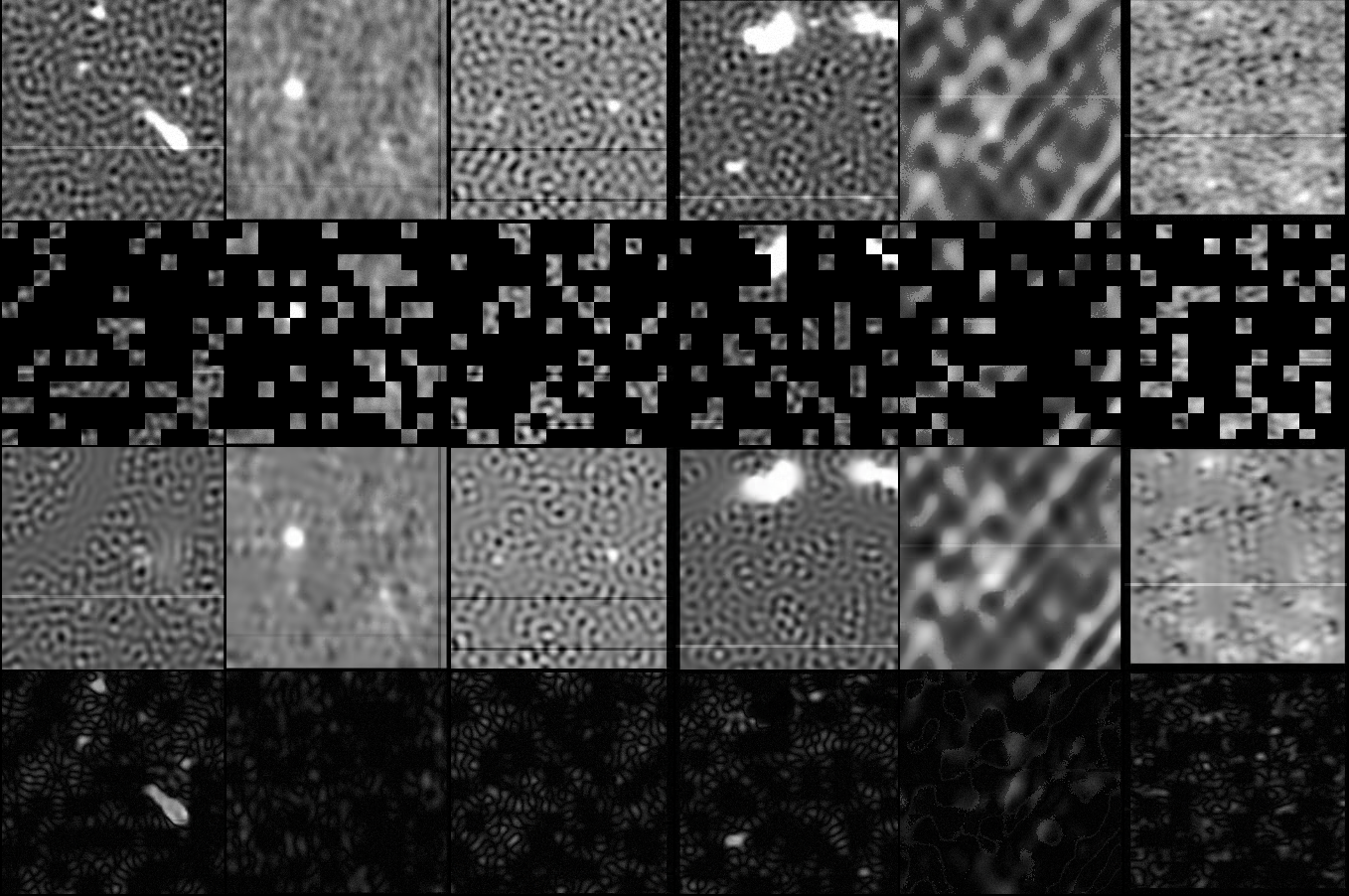}
\end{minipage}
\caption{Representative masked-reconstruction diagnostics. The left and right panels contain independent sets of six examples. Within each panel, rows show, from top to bottom, the original reconstruction view, the masked input presented to the model, the decoder reconstruction, and the absolute pixel residual between the original and reconstructed views. The reconstructions recover the principal source morphology and broad intensity structure, whereas residuals remain concentrated around compact peaks, sharp boundaries, and fine-scale image structure. Unlike Figure~\ref{fig:mae_views}, this figure assesses reconstruction behavior rather than view construction.}\label{fig:mae_recon_examples}
\end{figure*}

Given an input cutout, we sample one of two geometric regimes:
\begin{enumerate}
\item \textbf{Wide-field crop (probability $p_{\mathrm{global}}=0.2$):} a square random resized crop is taken directly from the
full cutout with scale range $[0.70,1.00]$ and resized to the model input size.
\item \textbf{ROI-aligned crop (probability $1-p_{\mathrm{global}}$):} an object-centric square region of interest (ROI) is
identified by proposing candidate anchors using two complementary heuristics: (i) a $16\times 16$ tile grid is searched for
high-significance peaks (top $K_{\mathrm{z}}=3$ candidates, minimum separation of two tiles, and a minimum z-score floor of
$2.5$), and (ii) the centers of the top $K_{\mathrm{mean}}=5$ tiles from an $8\times 8$ grid scored by mean absolute intensity
are retained as additional candidates. If mean-based candidates are available, we select between the two candidate sets with
probability $p_{\mathrm{mean}}=0.5$, and then sample uniformly from the chosen set. Candidate anchors are filtered by requiring
their local peak intensity to exceed $0.8$ times the global peak of the parent cutout. A square
ROI is then sampled around this anchor with side length fraction in $[0.55,0.80]$ of the cutout size, and a wide/global
view is sampled inside the ROI with scale range $[0.50,0.70]$ (resized to the model input size), with bounded resampling if an
uninformative crop is proposed.
\end{enumerate}
After the geometric step, we apply the mild augmentation regime in Table~\ref{tab:aug_params}, clamp to $[0,1]$, and then
apply model-dependent channel normalization (i.e., ImageNet normalization for pretrained backbones).
The anchor z-score is $(I_{\rm peak}-m_{\rm bg})/\max(1.4826\,\mathrm{MAD}_{\rm bg},10^{-6})$, where background statistics
are estimated from pixels below the 80th percentile of a sample of at most 8192 pixels in the standardized view.
Figure~\ref{fig:mae_views} illustrates representative reconstruction-branch views produced by this pipeline.
Figure~\ref{fig:mae_recon_examples} shows masked-reconstruction outputs for representative views. The decoder
recovers the dominant compact and extended emission despite the high masking fraction, while the residuals are
largest around bright, sharply bounded, or fine-scale structure. These examples provide a qualitative diagnostic
of the reconstruction objective rather than evidence of photometric recovery.

\subsubsection{Augmentation Operators and Parameters}
All augmentations operate after ZScale stretching to $[0,1]$. Table~\ref{tab:aug_params} summarizes the mild regime used in
the reconstruction branch and the strong regime used for the additional contrastive view.
\begin{itemize}
\item \textbf{Dihedral transforms:} random $k\times 90^\circ$ rotation with $k\in\{0,1,2,3\}$ and horizontal flip with probability $0.5$.
\item \textbf{Asinh contrast stretch:} random asinh remapping to compress bright structure and lift faint emission.
\item \textbf{Small affine jitter:} bounded translation and isotropic scale perturbation.
\item \textbf{Additive noise:} add pixelwise Gaussian noise with standard deviation $\sigma$.
\item \textbf{Banding artifacts:} random horizontal or vertical bands with additive offset $\Delta$.
\item \textbf{Flux/background perturbation:} multiplicative attenuation of lower-intensity pixels below a threshold derived from the view statistics.
\end{itemize}
Unless stated otherwise, operators are applied in the order listed above.

\begin{table*}[pos=tb]
\caption{Augmentation parameter settings used for view generation. ``Mild'' applies to the reconstruction branch and to the wide view in the contrastive objective;
``Strong'' applies to the additional contrastive view.}\label{tab:aug_params}
\centering
\footnotesize
\setlength{\tabcolsep}{6pt}
\begin{tabular*}{\tblwidth}{@{}@{\extracolsep{\fill}}p{0.26\textwidth}p{0.32\textwidth}p{0.32\textwidth}@{}}
\toprule
Operator & Mild setting & Strong setting \\
\midrule
Dihedral transforms & $k\times90^\circ$ rotation ($k\in\{0,1,2,3\}$) + horizontal flip ($p=0.5$) & Same as mild \\
Asinh stretch & $p_{\mathrm{asinh}}=0.35$, $\alpha\in[3,12]$ & $p_{\mathrm{asinh}}=0.50$, $\alpha\in[3,20]$ \\
Affine jitter & translation $\tau=0.03$, scale jitter $\delta=0.05$ & translation $\tau=0.05$, scale jitter $\delta=0.08$ \\
Gaussian noise & $\sigma=0.02$ & $\sigma=0.04$ \\
Banding & reconstruction branch: $B=2$; contrastive wide view: disabled ($B=0$); $\Delta\in[-0.3,0.3]$ & $B=3$; $\Delta\in[-0.3,0.3]$ \\
Flux/background perturbation & $p=0.5$, $T=\mu+0.3\sigma_x$, $f\in[0.7,0.95]$ & $p=0.5$, $T=\mu+0.3\sigma_x$, $f\in[0.5,0.9]$ \\
\bottomrule
\end{tabular*}
\end{table*}

\subsection{Reconstruction Loss Variants}
The reconstruction branch follows the ViT-MAE masked image modeling objective \citep{He2022MAE}. Let $x$ denote an input view, $m$ a binary mask
over patch locations, and $\hat{x}$ the model reconstruction (after unpatchifying decoder outputs). The default ViT-MAE training
loss is a masked-patch mean squared error (MSE), computed only over the masked patches:
\begin{equation}
\mathcal{L}_{\mathrm{MAE}}(x,\hat{x};m)=\frac{1}{|m|}\sum_{j\in m} \lVert \hat{x}_{j}-x_{j} \rVert_{2}^{2}.
\end{equation}
We treat this as our baseline, denote it by $\mathcal{L}_{\mathrm{L2}}$, and set its weight to $w_{\mathrm{L2}}=1.0$:
\begin{equation}
\mathcal{L}_{\mathrm{L2}} = w_{\mathrm{L2}}\,\mathcal{L}_{\mathrm{MAE}}.
\end{equation}
We then compare two explicit modifications that add an additional reconstruction regularizer computed in the same normalized
input space as $x$. In the following, $\langle\cdot\rangle$ denotes the mean over all pixels (and channels) of an image.

\subsubsection{Masked-Patch MSE + Global L1}
To encourage fidelity under an absolute error metric, we add a global (all-pixel) L1 term:
\begin{equation}
\mathcal{L}_{\mathrm{L2+L1}} = w_{\mathrm{L2}}\,\mathcal{L}_{\mathrm{MAE}} + w_{\mathrm{L1}}\,\mathcal{L}_{\mathrm{L1}},
\end{equation}
where $\mathcal{L}_{\mathrm{L1}}=\langle|\hat{x}-x|\rangle$.
We use $w_{\mathrm{L2}}=1.0$ and $w_{\mathrm{L1}}=0.1$.

\subsubsection{Masked-Patch MSE + Brightness-Weighted L1}
As a second variant, we replace the global L1 with a brightness-weighted L1 term that up-weights residuals in brighter regions.
Let $w(x)$ be a per-pixel weight proportional to the input brightness and normalized to have unit mean:
\begin{equation}
w(x)=\frac{|x|}{\langle |x| \rangle+\epsilon},
\end{equation}
with $\epsilon>0$ for numerical stability. The objective is
\begin{equation}
\mathcal{L}_{\mathrm{L2+BL1}} = w_{\mathrm{L2}}\,\mathcal{L}_{\mathrm{MAE}} + w_{\mathrm{BL1}}\,\mathcal{L}_{\mathrm{BL1}},
\end{equation}
where $\mathcal{L}_{\mathrm{BL1}}=\langle w(x)\,|\hat{x}-x|\rangle$ and $w(x)$ is normalized so that $\langle w(x)\rangle\approx 1$.
We use $w_{\mathrm{L2}}=1.0$ and $w_{\mathrm{BL1}}=0.1$.

\section{Contrastive Learning Objective}\label{phase2}
The contrastive objective is used in two settings: directly in the contrastive-only branch, and as the second stage of the
two-stage branch. It learns invariances across multiple augmented views of the same standardized
cutout. We apply the contrastive objective to the $\ell_2$-normalized projection-head outputs (Section~\ref{vit_backbone}), using
all other views in the (distributed) batch as negatives. We compare three contrastive loss variants: a standard InfoNCE
formulation, a ``soft'' hard-negative reweighting variant (soft-HCL), and a hard-negative objective (HCL) \citep{Robinson2021HardNegatives}.

\subsection{View Generation for Contrastive Learning}
The contrastive branch requires correlated views of the same cutout. An anchored multi-view strategy ties
all views to a common ROI, reducing false-positive pairs in sparse fields where independent random crops can contain
different sources or only background.

\begin{figure*}[tb]
\centering
\includegraphics[width=\textwidth,height=0.90\textheight,keepaspectratio]{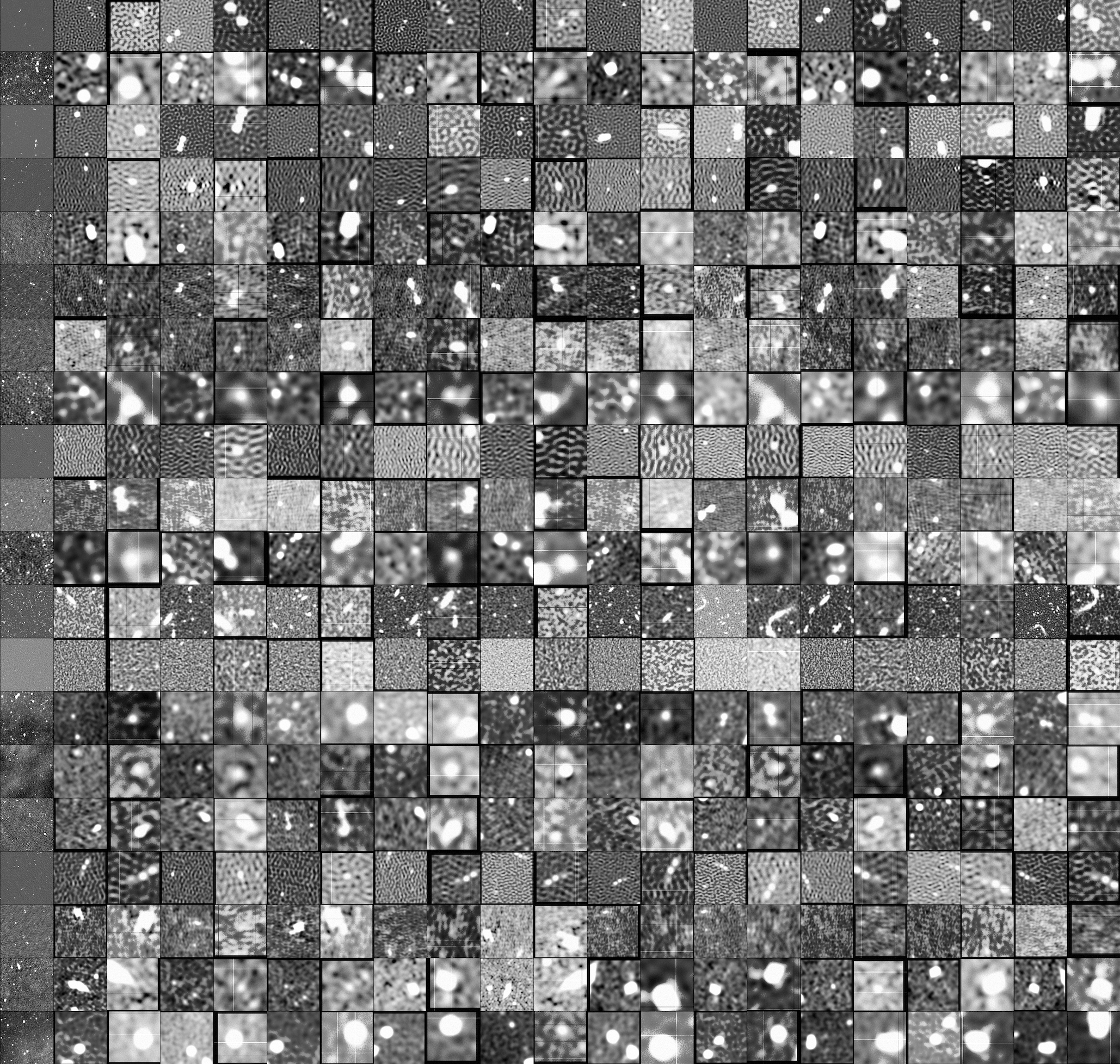}
\caption{Contrastive-branch multi-view examples produced by our on-the-fly augmenter. The first column shows the standardized
parent cutout. Each subsequent \emph{pair} of columns illustrates two correlated views sampled from the same cutout and
anchored to the same object-centric ROI (Section~\ref{phase2}): a wider/global view with mild, morphology-preserving
augmentations and an additional view with stronger corruptions. For training we sample exactly $V=2$ views per cutout per
forward pass, so each step uses one such pair per cutout; the multiple pairs shown here are independent draws for
visualization. In the contrastive loss, the two views of the same cutout form the positives, while all other views in the
(distributed) micro-batch act as negatives.}\label{fig:contrastive_views}
\end{figure*}

For each cutout, we first select an anchor region using the same anchor-selection scheme as in the reconstruction branch and sample an
object-centric ROI with side length fraction in $[0.55,0.80]$ of the cutout. In this work we generate two square views (resized
to the model input size) with fixed ordering:
\begin{itemize}
\item \textbf{Wide/global view:} sampled within the ROI using scale range $[0.50,0.70]$.
\item \textbf{Additional view:} sampled within the ROI using a narrower scale range $[0.20,0.35]$ and constrained to overlap
substantially with the wide view, requiring $\mathrm{IoA}\ge0.70$, ensuring shared content across views.
\end{itemize}
We define intersection-over-area (IoA) as
\begin{equation}
\mathrm{IoA}(a,b)=\frac{\mathrm{area}(a\cap b)}{\mathrm{area}(a)}.
\end{equation}
Here $a$ is the additional view and $b$ the wide view.
We use IoA rather than a symmetric IoU criterion so that each additional view lies mostly inside the wide view while the
wide view can retain broader context. Augmentations follow the same family as in the reconstruction branch, with the strong
regime applied to the additional view. Each view is clamped to $[0,1]$ and normalized using the same model-dependent policy.
Figure~\ref{fig:contrastive_views} shows representative contrastive pairs produced by this pipeline.

\subsubsection{Online Non-Empty Resampling}
Despite ROI anchoring, some proposed crops remain effectively empty. Rejection sampling is therefore applied at dataset level:
if the reconstruction view or any contrastive view fails a non-empty criterion, the pipeline resamples a different cutout
index. The wrapper uses a hard cap of 55 attempts (the configured five plus 50 fallback attempts), returning the
last retrieved sample if the non-empty criterion remains unsatisfied; if no sample was retrieved, it raises an error.
Empty views are therefore suppressed but not strictly excluded.

\subsection{Contrastive Loss Variants}\label{contrastive_losses}
Each training step operates on a batch of $B$ cutouts, generating $V$ views per cutout (in this work $V=2$) and producing an
embedding $z\in\mathbb{R}^{d}$ for each view via the encoder, mean pooling over patch tokens, and the projection head
(Section~\ref{vit_backbone}). Similarities are computed as dot products and scaled by a temperature $\tau$ that is linearly warmed from
$0.20$ to $0.10$ over the warmup portion of training (first 15\% of epochs) and then kept at $\tau=0.10$.
In our distributed implementation, the normalized embeddings are all-gathered across the 4 GPU ranks to form the key set. For
each local query, embeddings originating from the non-local ranks are used only as detached keys in the denominator, so gradients
are not propagated back through cross-device communication. This is a distributed implementation detail for InfoNCE-style
training rather than a SimSiam-style asymmetric stop-gradient design. In the ViT-MAE-based runs reported here, a single
forward pass uses 4 GPUs, per-device batch size 64, and 2 views per cutout, yielding 256 images and 512 keys in the
distributed micro-batch, i.e. 510 negatives per query. Gradient accumulation is set to 8 steps in this branch, giving an
effective global image batch of 2048, but the contrastive negative pool is determined only by the 512-key micro-batch seen in
one forward pass. InfoNCE, HCL, and Soft-HCL therefore share the same negative pool and differ only in how negatives are weighted.

\subsubsection{Baseline: InfoNCE}
We use a standard temperature-scaled InfoNCE contrastive loss (the NT-Xent form used in SimCLR \citep{Chen2020SimCLR}) as our
baseline.
Let $s_{ij}=z_i^\top z_j$ denote the similarity between a query $z_i$ and a key $z_j$. Let $P(i)$ be the set of positive keys
for query $i$. Let $A(i)$ denote the set of keys that appear in the denominator of the InfoNCE expression, i.e. all keys except
the query itself. Let $N(i)=A(i)\setminus P(i)$ denote the negatives for query $i$, meaning the denominator keys that are not
positives. Define the positive mass
\begin{equation}
p_i=\sum_{p\in P(i)}\exp(s_{ip}/\tau),
\end{equation}
and the per-negative masses $n_{ij}=\exp(s_{ij}/\tau)$ for $j\in N(i)$, with total negative mass
\begin{equation}
q_i=\sum_{j\in N(i)} n_{ij}.
\end{equation}
If $Q$ denotes the total number of query embeddings contributing to the loss, with both directions of each positive pair already
counted as separate queries, the baseline multi-positive InfoNCE loss can be written as
\begin{equation}
\mathcal{L}_{\mathrm{NCE}}=-\frac{1}{Q}\sum_{i=1}^{Q}\log\frac{p_i}{p_i+q_i}.
\end{equation}

\subsubsection{Hard Contrastive Learning (HCL)}
HCL \citep{Robinson2021HardNegatives} retains the InfoNCE log-ratio but gives greater denominator weight to
negatives more similar to the query.

Starting from the baseline denominator $p_i+q_i$, HCL replaces $q_i=\sum_{j\in N(i)}n_{ij}$ with a reweighted negative mass
$\tilde{q}^{\mathrm{HCL}}_i$ that emphasizes hard negatives. First compute the mean negative mass
$\bar{n}_i=\mathrm{mean}_{k\in N(i)}(n_{ik})$ and define importance weights
\begin{equation}
w^{\mathrm{HCL}}_{ij}=\beta\,\frac{n_{ij}}{\bar{n}_i},
\qquad j\in N(i),
\end{equation}
so that ``hard'' negatives (large $n_{ij}$) receive larger weights. For example, a negative with $n_{ij}=2\bar{n}_i$ receives
twice the weight of an average negative (for $\beta=1$). The resulting effective negative mass is
\begin{equation}
\tilde{q}^{\mathrm{HCL}}_i=\sum_{j\in N(i)} w^{\mathrm{HCL}}_{ij}\,n_{ij},
\end{equation}
equivalently, $\tilde{q}^{\mathrm{HCL}}_i=\beta\,\sum_{j\in N(i)} n_{ij}^{2}/\bar{n}_i$.

Finally, the HCL loss is
\begin{equation}\label{eq:hcl}
\mathcal{L}_{\mathrm{HCL}}=-\frac{1}{Q}\sum_{i=1}^{Q}\log\frac{p_i}{p_i+\tilde{q}^{\mathrm{HCL}}_i}.
\end{equation}
The original formulation also permits an optional debiasing correction controlled by $\tau_+$; in this work we use $\tau_+=0$
(disabled) and set $\beta=1.0$.
In this implementation $\beta$ scales the negative mass; it is not the concentration exponent of
\citet{Robinson2021HardNegatives}. The importance weights coincide with their exponent-one case at the fixed value used here.

\subsubsection{Soft Hard-Negative Reweighting (Soft-HCL)}
We introduce Soft-HCL as a ``softened'' alternative
to HCL in which negative weights are obtained from a mixture of uniform weights and a hardness distribution computed from
cosine similarities scaled by a separate temperature $\tau_h$. Using the same notation as above, we first define a hardness distribution over negatives
\begin{equation}
w^{\mathrm{hard}}_{ij}=\frac{\exp(s_{ij}/\tau_h)}{\sum_{k\in N(i)}\exp(s_{ik}/\tau_h)},\qquad j\in N(i),
\end{equation}
and a uniform distribution $w^{\mathrm{uni}}_{ij}=1/|N(i)|$. Note that $\tau_h$ is distinct from the contrastive temperature
$\tau$; it controls how concentrated the hardness distribution is, independent of the softmax temperature used in the loss.
We then form the mixture
\begin{equation}
w^{\mathrm{soft}}_{ij}=(1-\alpha)w^{\mathrm{uni}}_{ij}+\alpha w^{\mathrm{hard}}_{ij},
\end{equation}
scale the weights so that $\sum_{j\in N(i)}w^{\mathrm{soft}}_{ij}=|N(i)|$, and compute an effective negative mass
\begin{equation}
\tilde{q}^{\mathrm{soft}}_i=\sum_{j\in N(i)} w^{\mathrm{soft}}_{ij}\,n_{ij}.
\end{equation}
We then use the same loss form as in Eq.~\eqref{eq:hcl}, replacing $\tilde{q}^{\mathrm{HCL}}_i$ with $\tilde{q}^{\mathrm{soft}}_i$.
We use $\alpha=0.5$ and $\tau_h=0.15$.

The mixture aims to emphasize informative negatives without allowing a few extreme similarities to dominate.
In sparse radio cutouts, shared noise, artifacts, repeated morphology, or possible near-duplicates across mosaic
tiles could otherwise produce misleadingly hard negatives. This is a motivation for the loss, not a demonstrated
cause of the transfer results.

\begin{table}[pos=t]
\caption{Contrastive loss variants and hyperparameters used in the contrastive branch.}\label{tab:contrastive_params}
\centering
\setlength{\tabcolsep}{3pt}
\begin{tabular*}{\tblwidth}{@{}p{0.24\columnwidth}@{\extracolsep{\fill}}p{0.66\columnwidth}@{}}
\toprule
Variant & Parameters \\
\midrule
InfoNCE & temperature $\tau$: $0.20\rightarrow 0.10$\allowbreak\space (warmup 15\% epochs), then $\tau=0.10$ \\
HCL & $\tau$ as above; $\beta=1.0$, $\tau_+=0$ \\
Soft-HCL & $\tau$ as above; $\alpha=0.5$, $\tau_h=0.15$ \\
\bottomrule
\end{tabular*}
\end{table}

\section{Training Setup}\label{training_setup}
The main branches continue from public ImageNet-1K ViT-MAE weights, using ImageNet channel normalization.
Contrastive-only training starts from that initialization; the two-stage contrastive phase starts from its
architecture-matched reconstruction checkpoint. A separate DINOv2 initialization study is described below.

\subsection{Common Optimization Settings}\label{training_hparams}
Table~\ref{tab:train_hparams} specifies the shared optimization settings. All runs use seed 42 on
4$\times$ NVIDIA RTX 6000 Ada GPUs (48\,GB each). Data loading uses 8 workers per device, persistent workers,
prefetch factor 2, and pinned memory; metrics are logged every 100 optimizer steps and checkpoints saved every 5 epochs.

\subsection{Distributed Training and Normalization Details}
Data-parallel contrastive training uses the all-gather and detached non-local keys defined in
Section~\ref{phase2}; accumulation changes the optimizer batch, not the per-forward negative pool.
ViT-MAE projection-head BatchNorm statistics are synchronized across all four GPUs; the DINOv2 projector has no BatchNorm.

\subsection{Reconstruction-Branch Training}
Reconstruction uses the three losses and weights in Section~\ref{phase1}. Its 35-epoch schedule, masking,
and batch settings are given in Table~\ref{tab:train_hparams}; $256\times4\times2=2048$ images contribute to each update.

\subsection{Contrastive-Branch Training}
Contrastive training lasts 35 epochs, either alone or after the 35-epoch reconstruction stage.
Thus two-stage training totals 70 full-dataset epochs, compared with 35 for contrastive-only training. These budgets
refer to complete dataset passes, irrespective of any chunk-local epoch counters used by the data loader.
Each image supplies two unmasked views; together with retained cross-device keys, this raises memory use relative
to masked reconstruction. A per-device batch of 64 was the largest stable ViT-MAE setting on our hardware.
Eight accumulation steps give $64\times4\times8=2048$ images per update.
Temperature and clipping settings are retained in Table~\ref{tab:train_hparams}.

\subsection{DINOv2 Initialization Study}\label{dino_initialization}
DINOv2 trains a student network to match representations produced by an exponential-moving-average teacher across
augmented views, combining image-level self-distillation with masked patch prediction \citep{Oquab2023DINOv2}.
The baseline comparison includes DINOv2-Small (ViT-S/14: 12 blocks, 6 attention heads, $D=384$), DINOv2-Base
(ViT-B/14: 12 blocks, 12 attention heads, $D=768$), and DINOv2-Base(R), which adds four register tokens to the
Base architecture. The adaptation experiment is restricted to DINOv2-Base and DINOv2-Base(R), both trained with
the fixed contrastive-only HCL configuration. It assesses applicability of the adaptation to a second pretrained ViT
family rather than providing a fully architecture-matched MAE--DINO comparison. Repeating the full loss grid is
computationally prohibitive:
native DINOv2 inputs ($518\times518$, 1369 patch tokens before prefix tokens) increase runtime from roughly
8\,h to 100\,h relative to ViT-MAE ($224\times224$, 196 tokens) on our hardware.

The longer sequence requires a per-device batch of 20 and 26 accumulation steps, giving
$20\times4\times26=2080$ images per update, the nearest feasible value to the 2048-image target.
Despite similar optimizer batches, the four-GPU, two-view key set falls from 512 to 160 views and the
negatives per query from 510 to 158. This mismatch limits direct attribution of differences to initialization.

Table~\ref{tab:train_hparams} provides a compact summary of the default branch-specific settings.
\begin{table*}[pos=tb]
\caption{Default ViT-MAE-branch training settings (each stage lasts 35 epochs). DINOv2 uses the batch and resolution adjustments described in the text.}\label{tab:train_hparams}
\centering
\footnotesize
\setlength{\tabcolsep}{6pt}
\begin{tabular*}{\tblwidth}{@{}@{\extracolsep{\fill}}p{0.26\textwidth}p{0.34\textwidth}p{0.34\textwidth}@{}}
\toprule
Setting & Reconstruction branch & Contrastive branch \\
\midrule
Epochs & 35 & 35 \\
Patch size & 16 & 16 \\
Masking & mask ratio $0.75$ & unmasked encoder (no patch masking) \\
Views per cutout & 1 & $V=2$ \\
Per-device batch size & 256 & 64 \\
Gradient-accumulation steps & 2 & 8 \\
Effective global batch & \multicolumn{2}{p{0.68\textwidth}}{2048 images via gradient accumulation (before view replication)} \\
Optimizer & \multicolumn{2}{p{0.68\textwidth}}{AdamW (fused on CUDA), weight decay $0.05$, $(\beta_1,\beta_2)=(0.9,0.95)$} \\
Learning-rate schedule & \multicolumn{2}{p{0.68\textwidth}}{cosine decay with linear warmup (15\% of epochs), base LR $5\times10^{-4}$} \\
Precision & \multicolumn{2}{p{0.68\textwidth}}{bf16} \\
Gradient clipping & $\lVert g\rVert_2 \le 1.0$ & $\lVert g\rVert_2 \le 3.0$ \\
Temperature & n/a & $\tau: 0.20\rightarrow 0.10$ (warmup), then $\tau=0.10$ \\
\bottomrule
\end{tabular*}
\end{table*}

\section{Transfer Evaluation Protocol}\label{transfer_eval}

\subsection{Common Evaluation Setup}\label{eval_common}
\begin{enumerate}
\item \textbf{Preprocessing and resizing:} We use the common evaluation preprocessing in
Section~\ref{classification_data}, retaining each model family's input resolution.
\item \textbf{Augmentations:} During training (both linear probe and fine-tuning), we apply only simple dihedral (D4)
transforms: a random rotation by $k\times90^\circ$ with $k\in\{0,1,2,3\}$ and a horizontal flip with probability 0.5
(implemented with \texttt{torch.rot90} and \texttt{torch.flip}, so no interpolation is introduced). Evaluation uses no
stochastic augmentations.
\item \textbf{Input normalization:} Channel-wise normalization uses the mean and standard deviation stored with each
checkpoint's image processor, i.e. ImageNet mean/std for ImageNet-initialized checkpoints.
\item \textbf{Cross-validation:} Each dataset uses $K=3$ stratified folds, shuffled with seed 0:
two folds for training and one held out. Final-epoch metrics are summarized as mean $\pm$ standard deviation
across folds. We use neither a further validation split nor early stopping, preserving labeled training data
and the same optimization budget across variants. Class-weighted cross-entropy uses balanced weights
computed from dataset class frequencies.
\item \textbf{Metrics:} Macro-F1, our primary metric for these imbalanced benchmarks, weights classes equally;
Weighted-F1 weights them by support. Summed fold confusion matrices provide the classwise diagnostics
in Section~\ref{results_classwise}.
\item \textbf{Readout:} MAE-family models mean-pool patch tokens as in Section~\ref{vit_backbone};
DINOv2-family models use the final LayerNorm-normalized CLS representation defined in Section~\ref{vit_backbone}.
\item \textbf{Shared optimization:} LP and FT both use 20 epochs of AdamW \citep{Loshchilov2019AdamW}, weight decay $0.05$, cosine decay,
gradient clipping at $\lVert g\rVert_2\le1.0$, bf16 precision, four GPUs, and an effective global batch of 256 images.
\end{enumerate}

\subsection{Linear Probing}\label{linear_probe}
A strict linear probe learns only $W\in\mathbb{R}^{C\times D}$ and $b$ in $\ell=Wz+b$, where $z$ is the
representation defined above and $C$ the class count. All encoder parameters are frozen and the backbone stays
in evaluation mode. Additional classification-head normalization, dropout, and MLP layers are disabled;
frozen backbone-internal normalization remains part of the representation.

The probe learning rate is $5\times10^{-3}$ with warmup ratio 0.05; all other optimization settings are shared above.

\subsection{Fine-Tuning}\label{fine_tune}
Fine-tuning optimizes the encoder and classification head end-to-end. For MAE-family models, the head comprises
LayerNorm, optional dropout, and a linear layer, without BatchNorm. The backbone learning rate is
$5\times10^{-4}$, with a head multiplier of 10, warmup ratio 0.2, and layer-wise learning-rate decay of 0.65
toward earlier Transformer blocks. These MAE-style settings were selected during preliminary development by
monitoring optimization stability and training-metric behavior, without inspecting downstream test-set performance;
they were then fixed across variants rather than tuned separately for each model or dataset.
The learning rates, head multiplier, warmup, and layer-wise decay are also applied to DINOv2. Its head is a linear
classifier on the native normalized CLS token, rather than the additional MAE-family LayerNorm head. The shared
optimization recipe supports a controlled transfer comparison; the unstable DINOv2-Base MiraBest FT result is therefore
interpreted as performance under this common recipe rather than as an architecture-wide ranking.

\subsection{Release-Checkpoint Selection Criterion}\label{checkpoint_selection}
The exploratory grid is used to select one continued-pretraining release checkpoint before the repeated-seed evaluation.
Candidates comprise all ViT-MAE reconstruction-only, contrastive-only, and two-stage checkpoints in the reported grid,
with and without registers. DINOv2 checkpoints are a separate initialization study and are not eligible. Within each
dataset, we average the LP and FT mean Macro-F1 values for each candidate and then select the candidate that maximizes the
minimum of these three dataset-level averages. This gives equal status to the three benchmarks and favors comparatively
robust cross-dataset transfer rather than an individual benchmark maximum.
Using the displayed means, the selected L2+L1/Soft-HCL ($R=4$) checkpoint scores 0.6205, versus 0.6200 for
L2/Soft-HCL ($R=4$), a margin of only 0.05 percentage points. This is below the precision of the individual reported
means and provides an operational release choice, not evidence that this loss combination is superior. The separately
evaluated HCL-adapted DINOv2-Base(R) would score 0.6195 on the same displayed grid.

\subsection{Post-Selection Repeated-Seed Evaluation}\label{post_selection_eval}
Applying the criterion in Section~\ref{checkpoint_selection} fixes the STRADAViT checkpoint before formal repeated-seed
evaluation. The primary comparison evaluates it against the ViT-MAE initialization from which it was derived, holding the backbone
family and input resolution fixed. A secondary comparison evaluates the fixed HCL-adapted DINOv2-Base(R) checkpoint against
its off-the-shelf DINOv2-Base(R) initialization. Each model pair is evaluated under linear probing and fine-tuning with 15
matched random seeds, numbered 0--14. Within each dataset and transfer protocol, the same seed is assigned to both models in a pair,
controlling the random processes used for downstream-head initialization, minibatch ordering, and stochastic augmentation.

MiraBest uses the fixed MiraBest Confident Dataset-F partition of 729 training and 104 test images, which also permits the
literature comparison in Appendix~\ref{app:mirabest_standard_split}. For LoTSS DR2 and RGZ DR1, we fix the first fold of the
seed-0 stratified three-fold partition used in the exploratory evaluation: 5870/2935 training/test images for LoTSS DR2 and
65,594/32,797 for RGZ DR1. The held-out samples therefore remain unchanged
across all 15 repetitions. All other preprocessing, augmentation, class weighting, model readout, and optimization settings
are identical to Sections~\ref{eval_common}--\ref{fine_tune}. This design estimates sensitivity to downstream optimization
and augmentation randomness on fixed partitions.
The pretrained checkpoints are fixed throughout, and pretraining itself is not repeated across random seeds.
The 104 MiraBest test images are held out from downstream fitting within each repeat.

Macro-F1 is the pre-specified primary endpoint. For each model pair, dataset, and transfer protocol, we compute the seed-wise
paired difference, its pointwise 95\% Student-$t$ confidence interval, and a two-sided paired $t$-test with 14 degrees of
freedom. Differences are expressed in percentage points to indicate their magnitude on the metric scale. To control the
family-wise error rate, the six STRADAViT comparisons and the six DINOv2 comparisons are treated as separate families, with
the $p$-values in each family adjusted using the Holm procedure \citep{Holm1979}. A difference is treated as statistically supported only when its 95\% confidence
interval excludes zero and the Holm-adjusted $p$-value is below 0.05. Accuracy and test error are additionally retained for
the MiraBest literature comparison, but they are not primary endpoints in the cross-dataset analysis.
As a sensitivity check, we also compute two-sided Wilcoxon signed-rank tests by exact enumeration of signs,
using midranks for ties and omitting zero differences, with Holm correction within the same six-comparison families.
This check avoids a Gaussian model for the paired differences but retains the signed-rank test's symmetry assumption.

\section{Results and Analysis}\label{results_analysis}

Linear probing (LP) measures frozen-feature separability and is relevant to backbone reuse and limited-label settings;
fine-tuning (FT) measures end-to-end downstream performance. FT generally gives higher absolute Macro-F1, whereas
supervised adaptation can reduce differences among initializations. The two protocols are therefore interpreted
separately: a larger gain over the initialization under LP does not imply a higher absolute LP score than under FT.
Accuracy is additionally reported for comparison with prior MiraBest studies.

The three-fold grid is exploratory: variants share fold assignments and seeds, but fine-grained rankings are
descriptive. The selected-checkpoint comparison instead uses 15 paired downstream seeds
(Section~\ref{post_selection_eval}). Because the LoTSS DR2 and RGZ DR1 partitions also informed exploration,
this assesses seed stability conditional on selection, not independent held-out confirmation.

\subsection{ROI-Aware View-Generation Ablation}\label{results_roi_ablation}
Table~\ref{tab:roi_view_ablation} isolates ROI-aware cropping in the register-based Soft-HCL contrastive-only
configuration. The no-ROI control replaces ROI-aware selection with standard random paired views,
holding initialization, loss, optimizer, duration, and downstream protocol fixed. This ablation evaluates the
ROI crop-selection mechanism; the remaining augmentation operators are unchanged and are not individually ablated.
It does not enter release-checkpoint selection.

\begin{table*}[pos=tb]
\caption{Matched contrastive-only ablation of view-generation strategy. Values are mean Macro-F1 $\pm$ standard deviation over $K=3$ folds; the delta is computed as ROI-aware minus standard/no-ROI views, in percentage points, before rounding the displayed means to three decimals.}\label{tab:roi_view_ablation}
\centering
\footnotesize
\setlength{\tabcolsep}{5pt}
\begin{tabular*}{\tblwidth}{@{}@{\extracolsep{\fill}}l*{6}{c}@{}}
\toprule
View strategy & MiraBest LP & LoTSS DR2 LP & RGZ DR1 LP & MiraBest FT & LoTSS DR2 FT & RGZ DR1 FT \\
\midrule
Standard/no-ROI views & $0.659\pm0.011$ & $0.527\pm0.012$ & $0.635\pm0.004$ & $0.715\pm0.007$ & $0.651\pm0.014$ & $0.796\pm0.002$ \\
ROI-aware views & \best{$0.691\pm0.010$} & \best{$0.560\pm0.012$} & \best{$0.728\pm0.003$} & \best{$0.735\pm0.014$} & \best{$0.667\pm0.007$} & \best{$0.822\pm0.003$} \\
Delta, ROI-aware $-$ standard & $+3.16$ & $+3.32$ & $+9.34$ & $+1.98$ & $+1.63$ & $+2.57$ \\
\bottomrule
\end{tabular*}
\end{table*}

ROI-aware sampling has higher mean Macro-F1 in all six comparisons, most strongly for RGZ DR1 LP
($+9.34$ percentage points). FT differences narrow to $+1.63$--$+2.57$ points, consistent with supervised
adaptation closing part of the gap. These fold-level comparisons remain descriptive.

\subsection{Baselines (DINOv2, ViT-MAE)}\label{results_baselines}
Table~\ref{tab:baseline_transfer} compares public checkpoints without radio astronomy pretraining under
the common transfer protocol. The suffix (R) denotes register tokens.

\begin{table*}[pos=tb]
\caption{Baseline transfer under linear probing (LP) and full fine-tuning (FT). Entries are mean $\pm$ SD over the same three folds; comparisons are descriptive. Bold marks the highest mean Macro-F1 per dataset and protocol, with Weighted-F1 breaking ties. (R) denotes register tokens.}\label{tab:baseline_transfer}
\centering
\footnotesize
\setlength{\tabcolsep}{3pt}
\begin{tabular*}{\tblwidth}{@{}@{\extracolsep{\fill}}ll*{6}{c}@{}}
\toprule
Model & Protocol & \multicolumn{2}{c}{MiraBest} & \multicolumn{2}{c}{LoTSS DR2} & \multicolumn{2}{c}{RGZ DR1} \\
\cmidrule(lr){3-4}\cmidrule(lr){5-6}\cmidrule(lr){7-8}
 & & Macro-F1 & Weighted-F1 & Macro-F1 & Weighted-F1 & Macro-F1 & Weighted-F1 \\
\midrule
\multirow{2}{*}{DINOv2-Base} & LP & \best{$0.717\pm0.014$} & $0.718\pm0.014$ & \best{$0.569\pm0.005$} & $0.751\pm0.004$ & \best{$0.661\pm0.004$} & $0.871\pm0.001$ \\
 & FT & $0.461\pm0.195$ & $0.451\pm0.202$ & \best{$0.708\pm0.008$} & $0.831\pm0.002$ & $0.808\pm0.003$ & $0.916\pm0.001$ \\
\addlinespace[0.2em]
\multirow{2}{*}{DINOv2-Small} & LP & $0.713\pm0.018$ & $0.713\pm0.018$ & $0.544\pm0.014$ & $0.729\pm0.007$ & $0.648\pm0.002$ & $0.861\pm0.001$ \\
 & FT & $0.707\pm0.031$ & $0.708\pm0.031$ & $0.704\pm0.014$ & $0.828\pm0.004$ & $0.795\pm0.002$ & $0.912\pm0.001$ \\
\addlinespace[0.2em]
\multirow{2}{*}{DINOv2-Base(R)} & LP & $0.685\pm0.013$ & $0.686\pm0.013$ & $0.529\pm0.009$ & $0.716\pm0.003$ & $0.629\pm0.001$ & $0.862\pm0.001$ \\
 & FT & $0.704\pm0.102$ & $0.704\pm0.102$ & $0.708\pm0.017$ & $0.829\pm0.004$ & \best{$0.812\pm0.002$} & $0.917\pm0.001$ \\
\addlinespace[0.2em]
\multirow{2}{*}{ViT-MAE-Base} & LP & $0.644\pm0.020$ & $0.645\pm0.020$ & $0.489\pm0.005$ & $0.663\pm0.006$ & $0.586\pm0.001$ & $0.841\pm0.000$ \\
 & FT & \best{$0.724\pm0.002$} & $0.725\pm0.002$ & $0.696\pm0.015$ & $0.819\pm0.001$ & $0.805\pm0.003$ & $0.916\pm0.001$ \\
\bottomrule
\end{tabular*}
\end{table*}

DINOv2-Base has the highest LP mean Macro-F1 on all three datasets, with DINOv2-Small close behind.
FT rankings vary by dataset: ViT-MAE leads on MiraBest, DINOv2-Base on LoTSS DR2
(tied in Macro-F1 with Base(R), but higher in Weighted-F1), and Base(R) on RGZ DR1.

Non-register DINOv2-Base is unstable on MiraBest FT ($0.461\pm0.195$). We retain this result for transparency
but do not use it as the sole ranking reference; DINOv2 comparisons use the stable variant with the highest
mean under the same protocol. On LoTSS DR2 and RGZ DR1, much higher Weighted-F1 than Macro-F1 indicates
that strong aggregate performance can coexist with weaker minority-class behavior.

\subsection{Continued Pretraining Branches}\label{results_continued}
\subsubsection{Reconstruction-Only Branch}\label{results_continued_phase1}
Table~\ref{tab:continued_p1_transfer} compares the reconstruction losses from Section~\ref{phase1},
continuing from ViT-MAE with $R\in\{0,4\}$ and otherwise fixed settings.

\begin{table*}[pos=tb]
\caption{Reconstruction-only transfer from ViT-MAE under linear probing (LP) and full fine-tuning (FT). Entries are mean $\pm$ SD over the same three folds; comparisons are descriptive. Bold marks the highest mean Macro-F1 per dataset and protocol, within each register setting, with Weighted-F1 breaking ties.}\label{tab:continued_p1_transfer}
\centering
\footnotesize
\setlength{\tabcolsep}{3pt}
\begin{tabular*}{\tblwidth}{@{}@{\extracolsep{\fill}}ll*{6}{c}@{}}
\toprule
Loss & Protocol & \multicolumn{2}{c}{MiraBest} & \multicolumn{2}{c}{LoTSS DR2} & \multicolumn{2}{c}{RGZ DR1} \\
\cmidrule(lr){3-4}\cmidrule(lr){5-6}\cmidrule(lr){7-8}
 & & Macro-F1 & Weighted-F1 & Macro-F1 & Weighted-F1 & Macro-F1 & Weighted-F1 \\
\midrule
\multicolumn{8}{@{}l}{\textbf{No registers ($R=0$)}}\\
\addlinespace[0.2em]
\multirow{2}{*}{L2} & LP & $0.607\pm0.013$ & $0.607\pm0.013$ & $0.433\pm0.006$ & $0.606\pm0.005$ & $0.526\pm0.005$ & $0.811\pm0.002$ \\
 & FT & $0.708\pm0.011$ & $0.709\pm0.011$ & \best{$0.661\pm0.013$} & $0.791\pm0.003$ & \best{$0.795\pm0.005$} & $0.915\pm0.001$ \\
\addlinespace[0.2em]
\multirow{2}{*}{L2+L1} & LP & $0.609\pm0.006$ & $0.609\pm0.006$ & $0.436\pm0.013$ & $0.599\pm0.009$ & $0.525\pm0.001$ & $0.813\pm0.001$ \\
 & FT & \best{$0.713\pm0.018$} & $0.713\pm0.018$ & $0.657\pm0.017$ & $0.793\pm0.003$ & $0.795\pm0.005$ & $0.914\pm0.001$ \\
\addlinespace[0.2em]
\multirow{2}{*}{L2+BL1} & LP & \best{$0.609\pm0.012$} & $0.610\pm0.012$ & \best{$0.448\pm0.008$} & $0.611\pm0.008$ & \best{$0.526\pm0.002$} & $0.813\pm0.001$ \\
 & FT & $0.708\pm0.025$ & $0.709\pm0.026$ & $0.654\pm0.013$ & $0.788\pm0.003$ & $0.795\pm0.005$ & $0.914\pm0.001$ \\
\midrule
\multicolumn{8}{@{}l}{\textbf{With registers ($R=4$)}}\\
\addlinespace[0.2em]
\multirow{2}{*}{L2} & LP & \best{$0.612\pm0.012$} & $0.613\pm0.012$ & \best{$0.485\pm0.017$} & $0.659\pm0.014$ & \best{$0.565\pm0.001$} & $0.832\pm0.001$ \\
 & FT & \best{$0.717\pm0.014$} & $0.718\pm0.014$ & \best{$0.664\pm0.015$} & $0.792\pm0.006$ & \best{$0.793\pm0.004$} & $0.914\pm0.001$ \\
\addlinespace[0.2em]
\multirow{2}{*}{L2+L1} & LP & $0.605\pm0.009$ & $0.606\pm0.009$ & $0.479\pm0.009$ & $0.658\pm0.013$ & $0.562\pm0.004$ & $0.831\pm0.002$ \\
 & FT & $0.703\pm0.018$ & $0.703\pm0.018$ & $0.647\pm0.021$ & $0.785\pm0.010$ & $0.793\pm0.003$ & $0.913\pm0.000$\\
\addlinespace[0.2em]
\multirow{2}{*}{L2+BL1} & LP & $0.608\pm0.013$ & $0.610\pm0.013$ & $0.478\pm0.013$ & $0.660\pm0.010$ & $0.560\pm0.003$ & $0.830\pm0.001$ \\
 & FT & $0.697\pm0.007$ & $0.697\pm0.007$ & $0.655\pm0.026$ & $0.787\pm0.008$ & $0.791\pm0.006$ & $0.913\pm0.002$ \\
\bottomrule
\end{tabular*}
\end{table*}

Under LP, registers raise LoTSS DR2 and RGZ DR1 means. Loss differences are small: L2+BL1 leads without
registers after tie-breaking, while plain L2 leads with registers.

Under FT, register-associated differences shrink and loss choice has limited impact. Reconstruction-only
results occupy a narrow band, generally below the contrastive branches.
All reconstruction-only mean Macro-F1 values are also below the unadapted ViT-MAE baseline under both protocols.
Any benefit of a subsequent reconstruction warm start is therefore conditional on the contrastive stage; reconstruction
alone does not improve transfer in this grid.

\subsubsection{Contrastive-Only Branch}\label{results_continued_contrastive_only}
Table~\ref{tab:continued_contrastive_only_transfer} compares direct ViT-MAE adaptation with InfoNCE,
Soft-HCL, and HCL, without a preceding reconstruction stage.

\begin{table*}[pos=tb]
\caption{Contrastive-only transfer from ViT-MAE under linear probing (LP) and full fine-tuning (FT). Entries are mean $\pm$ SD over the same three folds; comparisons are descriptive. Bold marks the highest mean Macro-F1 per dataset and protocol, within each register setting, with Weighted-F1 breaking ties.}\label{tab:continued_contrastive_only_transfer}
\centering
\footnotesize
\setlength{\tabcolsep}{3pt}
\begin{tabular*}{\tblwidth}{@{}@{\extracolsep{\fill}}ll*{6}{c}@{}}
\toprule
Loss & Protocol & \multicolumn{2}{c}{MiraBest} & \multicolumn{2}{c}{LoTSS DR2} & \multicolumn{2}{c}{RGZ DR1} \\
\cmidrule(lr){3-4}\cmidrule(lr){5-6}\cmidrule(lr){7-8}
 & & Macro-F1 & Weighted-F1 & Macro-F1 & Weighted-F1 & Macro-F1 & Weighted-F1 \\
\midrule
\multicolumn{8}{@{}l}{\textbf{No registers ($R=0$)}}\\
\addlinespace[0.2em]
\multirow{2}{*}{InfoNCE} & LP & $0.676\pm0.003$ & $0.677\pm0.002$ & $0.538\pm0.010$ & $0.710\pm0.007$ & $0.699\pm0.001$ & $0.878\pm0.000$ \\
 & FT & \best{$0.726\pm0.005$} & $0.727\pm0.005$ & \best{$0.672\pm0.011$} & $0.798\pm0.004$ & $0.818\pm0.003$ & $0.921\pm0.001$ \\
\addlinespace[0.2em]
\multirow{2}{*}{Soft-HCL} & LP & \best{$0.692\pm0.012$} & $0.693\pm0.012$ & \best{$0.549\pm0.009$} & $0.719\pm0.003$ & $0.709\pm0.001$ & $0.883\pm0.001$ \\
 & FT & $0.719\pm0.005$ & $0.720\pm0.015$ & $0.660\pm0.002$ & $0.799\pm0.003$ & $0.824\pm0.003$ & $0.923\pm0.001$ \\
\addlinespace[0.2em]
\multirow{2}{*}{HCL} & LP & $0.686\pm0.010$ & $0.687\pm0.011$ & $0.548\pm0.008$ & $0.718\pm0.002$ & \best{$0.713\pm0.004$} & $0.886\pm0.000$ \\
 & FT & $0.723\pm0.008$ & $0.724\pm0.007$ & $0.661\pm0.005$ & $0.798\pm0.004$ & \best{$0.825\pm0.003$} & $0.924\pm0.001$ \\
\midrule
\multicolumn{8}{@{}l}{\textbf{With registers ($R=4$)}}\\
\addlinespace[0.2em]
\multirow{2}{*}{InfoNCE} & LP & $0.662\pm0.028$ & $0.663\pm0.029$ & $0.558\pm0.023$ & $0.730\pm0.010$ & $0.724\pm0.002$ & $0.893\pm0.000$ \\
 & FT & $0.726\pm0.010$ & $0.727\pm0.010$ & $0.666\pm0.011$ & $0.795\pm0.004$ & $0.819\pm0.002$ & $0.922\pm0.001$ \\
\addlinespace[0.2em]
\multirow{2}{*}{Soft-HCL} & LP & \best{$0.691\pm0.010$} & $0.692\pm0.009$ & $0.560\pm0.012$ & $0.727\pm0.008$ & \best{$0.728\pm0.003$} & $0.896\pm0.000$ \\
 & FT & \best{$0.735\pm0.014$} & $0.736\pm0.013$ & \best{$0.667\pm0.007$} & $0.798\pm0.003$ & $0.822\pm0.003$ & $0.923\pm0.002$ \\
\addlinespace[0.2em]
\multirow{2}{*}{HCL} & LP & $0.678\pm0.014$ & $0.679\pm0.015$ & \best{$0.560\pm0.007$} & $0.728\pm0.003$ & $0.727\pm0.003$ & $0.895\pm0.001$ \\
 & FT & $0.722\pm0.013$ & $0.723\pm0.013$ & $0.665\pm0.003$ & $0.797\pm0.003$ & \best{$0.826\pm0.001$} & $0.924\pm0.001$ \\
\bottomrule
\end{tabular*}
\end{table*}

Contrastive-only training yields higher means than reconstruction alone under both protocols.
Loss and register effects depend on the dataset: Soft-HCL leads several MiraBest and LoTSS DR2 settings,
while HCL generally leads RGZ DR1, except for the register-based LP setting where Soft-HCL is highest.

\subsubsection{DINOv2 Initialization Ablation}\label{results_dinov2_init_ablation}
Table~\ref{tab:dinov2_init_ablation} reports fixed-HCL adaptation of both DINOv2 initializations, using the
same datasets, fold assignments, and downstream protocol as the main grid.

\begin{table*}[pos=tb]
\caption{DINOv2 initialization ablation under the contrastive-only HCL recipe (mean $\pm$ std over the same $K=3$ fold assignments used elsewhere). The highest mean Macro-F1 per dataset within each transfer setting is highlighted in bold; ties are broken by Weighted-F1. Differences are interpreted descriptively rather than as statistically established rankings.}\label{tab:dinov2_init_ablation}
\centering
\scriptsize
\setlength{\tabcolsep}{4pt}
\begin{tabular*}{\tblwidth}{@{}@{\extracolsep{\fill}}ll*{3}{cc}@{}}
\toprule
Setting & Initialization & \multicolumn{2}{c}{MiraBest} & \multicolumn{2}{c}{LoTSS DR2} & \multicolumn{2}{c}{RGZ DR1} \\
\cmidrule(lr){3-4}\cmidrule(lr){5-6}\cmidrule(lr){7-8}
 &  & Macro-F1 & Weighted-F1 & Macro-F1 & Weighted-F1 & Macro-F1 & Weighted-F1 \\
\midrule
Linear Probe & DINOv2-Base & $0.683\pm0.010$ & $0.683\pm0.010$ & $0.551\pm0.010$ & $0.724\pm0.006$ & $0.735\pm0.004$ & $0.894\pm0.000$ \\
Linear Probe & DINOv2-Base(R) & \best{$0.686\pm0.018$} & $0.687\pm0.018$ & \best{$0.561\pm0.005$} & $0.731\pm0.003$ & \best{$0.739\pm0.004$} & $0.895\pm0.000$ \\
\midrule
Fine-Tuning & DINOv2-Base & $0.721\pm0.020$ & $0.722\pm0.020$ & $0.676\pm0.009$ & $0.816\pm0.004$ & $0.816\pm0.000$ & $0.919\pm0.000$ \\
Fine-Tuning & DINOv2-Base(R) & \best{$0.738\pm0.014$} & $0.738\pm0.014$ & \best{$0.678\pm0.015$} & $0.816\pm0.006$ & \best{$0.828\pm0.002$} & $0.922\pm0.002$ \\
\bottomrule
\end{tabular*}
\end{table*}

After fixed-HCL adaptation, DINOv2-Base(R) has higher mean Macro-F1 than the adapted non-register variant
on all three datasets under LP and FT, reversing their off-the-shelf LP ordering.
Relative to each variant's own baseline, however, the pattern differs by protocol: LP means rise on RGZ DR1
for both variants, but fall on MiraBest and LoTSS DR2 for non-register Base; Base(R) has a near-unchanged
MiraBest mean and a higher LoTSS mean. Under FT, both adapted variants have higher means on MiraBest and
RGZ DR1 but lower means on LoTSS DR2.
The observed means indicate that HCL-based continued pretraining can be applied to DINOv2. The three-fold experiment does not,
however, support an inferential ranking of initialization families, particularly because their negative-pool sizes differ.
The repeated-seed comparison of DINOv2-Base(R) against its own initialization is reported in
Table~\ref{tab:selected_significance}.

\subsubsection{Two-Stage Branch}\label{results_continued_phase12}
Tables~\ref{tab:continued_p12_lp} and \ref{tab:continued_p12_ft} compare contrastive losses within each
reconstruction-loss and register setting of the two-stage branch.

\begin{table*}[pos=tb]
\caption{Continued pretraining (two-stage branch): linear-probe results (mean $\pm$ std over $K=3$ folds). The contrastive stage uses an unmasked encoder. The highest mean Macro-F1 per dataset within each register setting is highlighted in bold; ties are broken by Weighted-F1.}\label{tab:continued_p12_lp}
\centering
\scriptsize
\setlength{\tabcolsep}{3pt}
\begin{tabular*}{\tblwidth}{@{}@{\extracolsep{\fill}}ll*{3}{cc}@{}}
\toprule
Reconstruction loss & Contrastive loss & \multicolumn{2}{c}{MiraBest} & \multicolumn{2}{c}{LoTSS DR2} & \multicolumn{2}{c}{RGZ DR1} \\
\cmidrule(lr){3-4}\cmidrule(lr){5-6}\cmidrule(lr){7-8}
 &  & Macro-F1 & Weighted-F1 & Macro-F1 & Weighted-F1 & Macro-F1 & Weighted-F1 \\
\midrule
\multicolumn{8}{@{}l}{\textbf{No registers ($R=0$)}}\\
\addlinespace[0.2em]
\multirow{3}{*}{L2} & InfoNCE & $0.670\pm0.010$ & $0.671\pm0.010$ & $0.529\pm0.010$ & $0.705\pm0.006$ & $0.693\pm0.003$ & $0.877\pm0.001$ \\
 & Soft-HCL & $0.677\pm0.007$ & $0.678\pm0.007$ & $0.543\pm0.010$ & $0.715\pm0.004$ & $0.704\pm0.002$ & $0.881\pm0.000$ \\
 & HCL & $0.689\pm0.010$ & $0.690\pm0.011$ & $0.545\pm0.012$ & $0.714\pm0.006$ & $0.706\pm0.003$ & $0.884\pm0.001$ \\
\cmidrule(lr){1-8}
\multirow{3}{*}{L2+L1} & InfoNCE & $0.664\pm0.013$ & $0.665\pm0.013$ & $0.528\pm0.010$ & $0.701\pm0.006$ & $0.693\pm0.004$ & $0.876\pm0.001$ \\
 & Soft-HCL & $0.678\pm0.004$ & $0.679\pm0.004$ & \best{$0.564\pm0.007$} & $0.739\pm0.005$ & $0.701\pm0.002$ & $0.881\pm0.000$ \\
 & HCL & $0.678\pm0.008$ & $0.679\pm0.007$ & $0.547\pm0.015$ & $0.717\pm0.007$ & $0.707\pm0.005$ & $0.884\pm0.001$ \\
\cmidrule(lr){1-8}
\multirow{3}{*}{L2+BL1} & InfoNCE & $0.671\pm0.015$ & $0.673\pm0.016$ & $0.535\pm0.012$ & $0.706\pm0.006$ & $0.690\pm0.003$ & $0.876\pm0.000$ \\
 & Soft-HCL & $0.680\pm0.001$ & $0.681\pm0.001$ & $0.550\pm0.014$ & $0.722\pm0.008$ & $0.706\pm0.002$ & $0.883\pm0.000$ \\
 & HCL & \best{$0.697\pm0.013$} & $0.698\pm0.014$ & $0.552\pm0.011$ & $0.717\pm0.004$ & \best{$0.708\pm0.003$} & $0.884\pm0.001$ \\
\midrule
\multicolumn{8}{@{}l}{\textbf{With registers ($R=4$)}}\\
\addlinespace[0.2em]
\multirow{3}{*}{L2} & InfoNCE & $0.671\pm0.027$ & $0.671\pm0.027$ & $0.566\pm0.009$ & $0.735\pm0.004$ & $0.722\pm0.002$ & $0.892\pm0.001$ \\
 & Soft-HCL & $0.672\pm0.011$ & $0.673\pm0.012$ & \best{$0.575\pm0.005$} & $0.738\pm0.005$ & $0.727\pm0.003$ & $0.893\pm0.000$ \\
 & HCL & $0.693\pm0.001$ & $0.694\pm0.001$ & $0.571\pm0.009$ & $0.743\pm0.006$ & $0.731\pm0.004$ & $0.894\pm0.001$ \\
\cmidrule(lr){1-8}
\multirow{3}{*}{L2+L1} & InfoNCE & $0.675\pm0.017$ & $0.675\pm0.018$ & $0.566\pm0.014$ & $0.731\pm0.005$ & $0.718\pm0.002$ & $0.890\pm0.001$ \\
 & Soft-HCL & $0.674\pm0.023$ & $0.675\pm0.023$ & $0.564\pm0.007$ & $0.739\pm0.005$ & \best{$0.732\pm0.002$} & $0.894\pm0.002$ \\
 & HCL & \best{$0.699\pm0.016$} & $0.700\pm0.016$ & $0.565\pm0.012$ & $0.738\pm0.007$ & $0.730\pm0.004$ & $0.894\pm0.001$ \\
\cmidrule(lr){1-8}
\multirow{3}{*}{L2+BL1} & InfoNCE & $0.672\pm0.019$ & $0.673\pm0.020$ & $0.559\pm0.012$ & $0.730\pm0.004$ & $0.719\pm0.003$ & $0.891\pm0.001$ \\
 & Soft-HCL & $0.680\pm0.010$ & $0.680\pm0.010$ & $0.558\pm0.009$ & $0.730\pm0.003$ & $0.729\pm0.002$ & $0.894\pm0.001$ \\
 & HCL & $0.668\pm0.004$ & $0.669\pm0.005$ & $0.565\pm0.013$ & $0.737\pm0.007$ & $0.730\pm0.001$ & $0.894\pm0.001$ \\
\bottomrule
\end{tabular*}
\end{table*}

\begin{table*}[pos=tb]
\caption{Continued pretraining (two-stage branch): full fine-tuning results (mean $\pm$ std over $K=3$ folds). The contrastive stage uses an unmasked encoder. The highest mean Macro-F1 per dataset within each register setting is highlighted in bold; ties are broken by Weighted-F1.}\label{tab:continued_p12_ft}
\centering
\scriptsize
\setlength{\tabcolsep}{3pt}
\begin{tabular*}{\tblwidth}{@{}@{\extracolsep{\fill}}ll*{3}{cc}@{}}
\toprule
Reconstruction loss & Contrastive loss & \multicolumn{2}{c}{MiraBest} & \multicolumn{2}{c}{LoTSS DR2} & \multicolumn{2}{c}{RGZ DR1} \\
\cmidrule(lr){3-4}\cmidrule(lr){5-6}\cmidrule(lr){7-8}
 &  & Macro-F1 & Weighted-F1 & Macro-F1 & Weighted-F1 & Macro-F1 & Weighted-F1 \\
\midrule
\multicolumn{8}{@{}l}{\textbf{No registers ($R=0$)}}\\
\addlinespace[0.2em]
\multirow{3}{*}{L2} & InfoNCE & $0.714\pm0.002$ & $0.715\pm0.015$ & $0.664\pm0.013$ & $0.793\pm0.004$ & $0.824\pm0.003$ & $0.922\pm0.001$ \\
 & Soft-HCL & $0.732\pm0.007$ & $0.733\pm0.007$ & $0.661\pm0.007$ & $0.790\pm0.004$ & $0.822\pm0.003$ & $0.923\pm0.001$ \\
 & HCL & $0.727\pm0.003$ & $0.728\pm0.003$ & $0.654\pm0.008$ & $0.793\pm0.004$ & $0.824\pm0.004$ & $0.923\pm0.002$ \\
\cmidrule(lr){1-8}
\multirow{3}{*}{L2+L1} & InfoNCE & $0.715\pm0.009$ & $0.716\pm0.009$ & $0.655\pm0.006$ & $0.790\pm0.003$ & $0.821\pm0.003$ & $0.922\pm0.001$ \\
 & Soft-HCL & $0.723\pm0.018$ & $0.724\pm0.018$ & $0.654\pm0.003$ & $0.791\pm0.004$ & $0.821\pm0.002$ & $0.923\pm0.001$ \\
 & HCL & $0.719\pm0.001$ & $0.720\pm0.000$ & $0.656\pm0.006$ & $0.791\pm0.005$ & \best{$0.825\pm0.001$} & $0.924\pm0.001$ \\
\cmidrule(lr){1-8}
\multirow{3}{*}{L2+BL1} & InfoNCE & $0.725\pm0.014$ & $0.726\pm0.015$ & $0.649\pm0.004$ & $0.790\pm0.005$ & $0.821\pm0.003$ & $0.922\pm0.001$ \\
 & Soft-HCL & $0.725\pm0.011$ & $0.726\pm0.011$ & \best{$0.664\pm0.004$} & $0.797\pm0.002$ & $0.823\pm0.002$ & $0.922\pm0.001$ \\
 & HCL & \best{$0.733\pm0.011$} & $0.733\pm0.010$ & $0.641\pm0.011$ & $0.794\pm0.002$ & $0.821\pm0.004$ & $0.923\pm0.002$ \\
\midrule
\multicolumn{8}{@{}l}{\textbf{With registers ($R=4$)}}\\
\addlinespace[0.2em]
\multirow{3}{*}{L2} & InfoNCE & $0.726\pm0.006$ & $0.727\pm0.006$ & $0.667\pm0.013$ & $0.795\pm0.007$ & $0.817\pm0.002$ & $0.921\pm0.001$ \\
 & Soft-HCL & \best{$0.736\pm0.014$} & $0.737\pm0.014$ & $0.665\pm0.005$ & $0.799\pm0.004$ & $0.824\pm0.002$ & $0.924\pm0.001$ \\
 & HCL & $0.730\pm0.009$ & $0.731\pm0.009$ & $0.665\pm0.004$ & $0.803\pm0.001$ & $0.824\pm0.004$ & $0.924\pm0.001$ \\
\cmidrule(lr){1-8}
\multirow{3}{*}{L2+L1} & InfoNCE & $0.714\pm0.015$ & $0.715\pm0.015$ & $0.649\pm0.006$ & $0.792\pm0.004$ & $0.820\pm0.001$ & $0.921\pm0.001$ \\
 & Soft-HCL & $0.725\pm0.004$ & $0.726\pm0.011$ & \best{$0.677\pm0.014$} & $0.803\pm0.008$ & \best{$0.825\pm0.001$} & $0.924\pm0.001$ \\
 & HCL & $0.721\pm0.007$ & $0.722\pm0.007$ & $0.664\pm0.006$ & $0.801\pm0.003$ & $0.822\pm0.006$ & $0.923\pm0.001$ \\
\cmidrule(lr){1-8}
\multirow{3}{*}{L2+BL1} & InfoNCE & $0.734\pm0.010$ & $0.734\pm0.010$ & $0.659\pm0.015$ & $0.793\pm0.007$ & $0.819\pm0.002$ & $0.921\pm0.001$ \\
 & Soft-HCL & $0.732\pm0.008$ & $0.732\pm0.008$ & $0.654\pm0.010$ & $0.795\pm0.005$ & $0.822\pm0.001$ & $0.922\pm0.001$ \\
 & HCL & $0.727\pm0.009$ & $0.728\pm0.009$ & $0.655\pm0.006$ & $0.798\pm0.002$ & $0.822\pm0.001$ & $0.923\pm0.001$ \\
\bottomrule
\end{tabular*}
\end{table*}

No single reconstruction--contrastive loss combination has the highest mean across all datasets and protocols. Register
tokens have their clearest
effect under LP on LoTSS DR2 and RGZ DR1, whereas the FT results occupy a narrower range and favor different loss
combinations by dataset. This variation motivates the cross-dataset selection criterion used below rather than selection
from any single table maximum.

Table~\ref{tab:contrastive_vs_twostage} compares architecture- and loss-matched ViT-MAE branches along the selected
release pathway. It evaluates the addition of the L2+L1 reconstruction warm start to a register-based Soft-HCL model.
DINOv2 is excluded because its token count and micro-batch differ.

\begin{table*}[pos=tb]
\caption{Architecture- and loss-matched comparison of register-based ($R=4$) Soft-HCL contrastive-only training with
L2+L1 reconstruction followed by Soft-HCL. Values are mean Macro-F1 over the same three exploratory folds. Delta is
$(\text{two-stage}-\text{contrastive-only})\times100$ percentage points and is interpreted descriptively.}
\label{tab:contrastive_vs_twostage}
\centering
\footnotesize
\setlength{\tabcolsep}{5pt}
\begin{tabular*}{\tblwidth}{@{}@{\extracolsep{\fill}}llccc@{}}
\toprule
Protocol & Branch & MiraBest & LoTSS DR2 & RGZ DR1 \\
\midrule
\multirow{3}{*}{LP} & Contrastive-only, Soft-HCL & $0.691$ & $0.560$ & $0.728$ \\
 & Two-stage, L2+L1 $\rightarrow$ Soft-HCL & $0.674$ & $0.564$ & $0.732$ \\
 & Delta (pp) & $-1.7$ & $+0.4$ & $+0.4$ \\
\midrule
\multirow{3}{*}{FT} & Contrastive-only, Soft-HCL & $0.735$ & $0.667$ & $0.822$ \\
 & Two-stage, L2+L1 $\rightarrow$ Soft-HCL & $0.725$ & $0.677$ & $0.825$ \\
 & Delta (pp) & $-1.0$ & $+1.0$ & $+0.3$ \\
\bottomrule
\end{tabular*}
\end{table*}

The reconstruction warm start raises mean Macro-F1 in four of the six comparisons, improving both LP and FT on
LoTSS DR2 and RGZ DR1; the largest gain is on LoTSS DR2 FT ($+1.0$ pp). MiraBest declines under both protocols.
The selected two-stage model is preferred under the cross-dataset criterion, while contrastive-only training remains
competitive at lower pretraining cost. With 35 epochs per stage, this comparison assesses the complete two-stage
recipe against 35-epoch contrastive-only training, rather than an equal-budget substitution of objectives.

\subsection{Selected STRADAViT Model: Exploratory Summary}\label{results_winner_analysis}
\begin{table*}[pos=tb]
\caption{Mean Macro-F1 and percentage-point (pp) changes for the fixed release checkpoint (L2+L1 reconstruction, Soft-HCL contrastive refinement, and four register tokens), relative to the ViT-MAE starting point and the highest-mean stable DINOv2 comparator. The DINOv2 references are Base for all LP cells, Small for MiraBest FT, Base for LoTSS DR2 FT, and Base(R) for RGZ DR1 FT. Each delta is computed as $(\text{release-checkpoint mean Macro-F1} - \text{reference mean Macro-F1})\times 100$. Positive values indicate higher mean Macro-F1.}\label{tab:delta_summary}
\centering
\scriptsize
\setlength{\tabcolsep}{4pt}
\begin{tabular*}{\tblwidth}{@{}@{\extracolsep{\fill}}ll*{3}{c}@{}}
\toprule
Setting & Model / comparison & MiraBest & LoTSS DR2 & RGZ DR1 \\
\midrule
\multicolumn{5}{@{}l}{\textbf{Linear Probe}}\\
\addlinespace[0.2em]
 & Release checkpoint & $0.674$ & $0.564$ & $0.732$ \\
 & Delta vs ViT-MAE (pp) & $+3.0$ & $+7.5$ & $+14.6$ \\
 & Delta vs highest-mean stable DINOv2 (pp) & $-4.3$ & $-0.5$ & $+7.1$ \\
\addlinespace[0.3em]
\midrule
\multicolumn{5}{@{}l}{\textbf{Fine-Tuning}}\\
\addlinespace[0.2em]
 & Release checkpoint & $0.725$ & $0.677$ & $0.825$ \\
 & Delta vs ViT-MAE (pp) & $+0.1$ & $-1.9$ & $+2.0$ \\
 & Delta vs highest-mean stable DINOv2 (pp) & $+1.8$ & $-3.1$ & $+1.3$ \\
\bottomrule
\end{tabular*}
\end{table*}

Applying the selection rule in Section~\ref{checkpoint_selection} selects the register-based L2+L1/Soft-HCL checkpoint.
It has a limiting dataset-level average of 0.6205 and uses 196 patch tokens per view, versus 1369 for DINOv2.

Table~\ref{tab:delta_summary} reports this fixed checkpoint throughout. For example, its MiraBest LP
difference from ViT-MAE is $(0.674-0.644)\times100=+3.0$ pp. Unlike the branch maxima, these comparisons
describe the same selected encoder across datasets and protocols.

\subsection{Repeated-Seed Analysis of Fixed Checkpoints}\label{results_significance}
Table~\ref{tab:selected_significance} reports the fixed STRADAViT--ViT-MAE and adapted--baseline DINOv2-Base(R)
comparisons over 15 paired downstream seeds. Inference follows the paired differences, confidence intervals, and
Holm-adjusted $p$-values under Section~\ref{post_selection_eval}, not rankings in the exploratory grid.

\begin{table*}[tb]
\caption{Repeated-seed comparisons of the fixed STRADAViT checkpoint with its ViT-MAE initialization and the HCL-adapted
DINOv2-Base(R) checkpoint with its off-the-shelf initialization. Macro-F1 values are mean $\pm$ standard deviation over
15 matched seeds on fixed train--test partitions. Delta is adapted model minus reference model in percentage points (pp);
CI denotes its pointwise 95\% confidence interval, also in pp.
MiraBest uses the standard Confident $729/104$ split; LoTSS DR2 and RGZ DR1 use the first exploratory fold.
$p$ is from a two-sided paired $t$-test, and $p_{\mathrm{Holm}}$ is adjusted separately across the six comparisons within
each model pair.}
\label{tab:selected_significance}
\centering
\footnotesize
\setlength{\tabcolsep}{4pt}
\begin{tabular*}{\tblwidth}{@{}@{\extracolsep{\fill}}llccccc@{}}
\toprule
Dataset & Protocol & Reference & Adapted model & $\Delta$ [95\% CI] (pp) & $p$ & $p_{\mathrm{Holm}}$ \\
\midrule
\multicolumn{7}{@{}l}{\textbf{STRADAViT versus ViT-MAE initialization}}\\
\addlinespace[0.2em]
MiraBest & LP & $0.660\pm0.007$ & $0.838\pm0.023$ & $+17.78\;[16.52,19.04]$ & $3.77\times10^{-14}$ & $1.51\times10^{-13}$ \\
MiraBest & FT & $0.825\pm0.014$ & $0.847\pm0.010$ & $+2.16\;[1.37,2.96]$ & $4.17\times10^{-5}$ & $4.17\times10^{-5}$ \\
LoTSS DR2 & LP & $0.495\pm0.003$ & $0.572\pm0.007$ & $+7.77\;[7.33,8.21]$ & $1.71\times10^{-15}$ & $8.54\times10^{-15}$ \\
LoTSS DR2 & FT & $0.700\pm0.009$ & $0.674\pm0.006$ & $-2.61\;[-3.29,-1.93]$ & $9.80\times10^{-7}$ & $1.96\times10^{-6}$ \\
RGZ DR1 & LP & $0.586\pm0.002$ & $0.733\pm0.002$ & $+14.65\;[14.52,14.78]$ & $1.13\times10^{-26}$ & $6.80\times10^{-26}$ \\
RGZ DR1 & FT & $0.809\pm0.004$ & $0.823\pm0.003$ & $+1.35\;[1.13,1.57]$ & $2.89\times10^{-9}$ & $8.67\times10^{-9}$ \\
\midrule
\multicolumn{7}{@{}l}{\textbf{HCL-adapted DINOv2-Base(R) versus off-the-shelf DINOv2-Base(R)}}\\
\addlinespace[0.2em]
MiraBest & LP & $0.840\pm0.011$ & $0.827\pm0.014$ & $-1.28\;[-2.27,-0.30]$ & $0.014$ & $0.029$ \\
MiraBest & FT & $0.870\pm0.018$ & $0.860\pm0.018$ & $-1.05\;[-2.58,+0.48]$ & $0.164$ & $0.164$ \\
LoTSS DR2 & LP & $0.571\pm0.004$ & $0.577\pm0.004$ & $+0.56\;[+0.31,+0.81]$ & $3.05\times10^{-4}$ & $9.14\times10^{-4}$ \\
LoTSS DR2 & FT & $0.725\pm0.008$ & $0.685\pm0.006$ & $-4.05\;[-4.68,-3.42]$ & $1.51\times10^{-9}$ & $7.56\times10^{-9}$ \\
RGZ DR1 & LP & $0.649\pm0.002$ & $0.735\pm0.001$ & $+8.61\;[+8.51,+8.71]$ & $4.62\times10^{-25}$ & $2.77\times10^{-24}$ \\
RGZ DR1 & FT & $0.822\pm0.002$ & $0.832\pm0.003$ & $+1.03\;[+0.84,+1.22]$ & $1.07\times10^{-8}$ & $4.29\times10^{-8}$ \\
\bottomrule
\end{tabular*}
\end{table*}

Under linear probing, STRADAViT exceeds ViT-MAE for every seed on every dataset. The paired gains are 17.78 pp on MiraBest,
7.77 pp on LoTSS DR2, and 14.65 pp on RGZ DR1, with confidence intervals well above zero. The pattern supports improved
linear separability of the frozen representations across all three benchmarks, beyond variation due to downstream training
seeds on these partitions.

Fine-tuning raises absolute Macro-F1 for both models but reduces the advantage of continued pretraining. STRADAViT improves
on MiraBest by 2.16 pp (13 of 15 positive paired differences) and on RGZ DR1 by 1.35 pp (15 of 15). On LoTSS DR2,
however, it falls below ViT-MAE by 2.61 pp in the mean and in every seed. All six comparisons remain significant after Holm
correction, including this deficit. The LoTSS result is therefore a reproducible limitation under the tested fine-tuning
protocol, rather than a chance ordering of fold means. It also shows that better frozen features need not yield better
end-to-end performance. The interpretation is therefore restricted to measured transfer behavior and does not attribute
the smaller relative FT gains specifically to overfitting.

For DINOv2, the direction of the effect again depends on dataset and protocol. HCL adaptation raises Macro-F1 on LoTSS DR2 LP by 0.56 pp
and on RGZ DR1 by 8.61 pp under LP and 1.03 pp under FT. It lowers MiraBest LP by 1.28 pp and LoTSS DR2 FT by 4.05 pp;
these five differences remain statistically supported after Holm correction. The MiraBest FT difference is $-1.05$ pp,
but its confidence interval includes zero and its Holm-adjusted $p$-value is 0.164. The repeated-seed results therefore
support applicability of the adaptation procedure to DINOv2, but not a general performance benefit.

Because all three evaluation pools informed exploration, these tests quantify downstream-seed variability conditional
on checkpoint selection, rather than independent confirmation on new samples.
The distinct MiraBest protocols are detailed in Appendix~\ref{app:mirabest_standard_split}.
In particular, the MiraBest exploratory evaluation mixes FIRST and NVSS images, whereas the repeated-seed test uses
FIRST only. Its selected-model LP mean changes from 0.674 to 0.838, compared with 0.644 to 0.660 for ViT-MAE and
0.685 to 0.840 for off-the-shelf DINOv2-Base(R). These are different image populations and partitions, not repeated
measurements of an unchanged benchmark. The comparison therefore reflects the combined change in survey resolution,
sample composition, and partition size.

Off-the-shelf DINOv2-Base(R) has higher mean Macro-F1 than STRADAViT on MiraBest LP and FT and on LoTSS DR2 FT;
the HCL-adapted variant also leads on RGZ DR1 under both protocols. These cross-family orderings are descriptive,
since the paired tests compare each adapted model with its own initialization. The ViT-MAE release provides gains
over that initialization at 196 rather than 1369 patch tokens per view; the roughly 8\,h versus 100\,h contrastive-stage
runtimes motivate this lower-cost choice, without establishing an overall performance optimum or a matched inference-time benchmark.

The signed-rank sensitivity check retains all six STRADAViT decisions (Holm-adjusted $p\le4.28\times10^{-4}$).
For DINOv2 it retains five differences (adjusted $p\le0.0135$), while MiraBest FT remains inconclusive ($p=0.1514$).

\subsection{Classwise Analysis of the Selected Model}\label{results_classwise}

\begin{figure*}[tp]
\centering
\includegraphics[width=\textwidth]{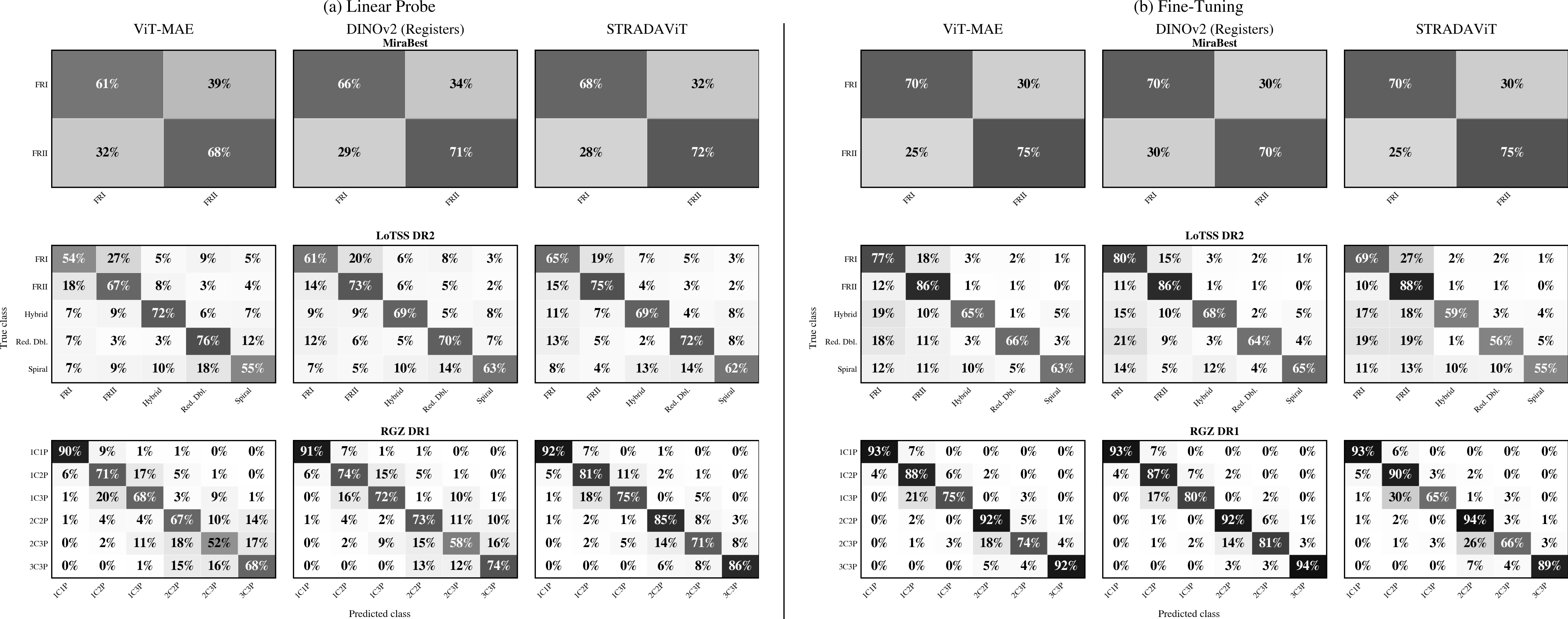}
\caption{Aggregate recall-form confusion matrices (\%) for MiraBest, LoTSS DR2, and RGZ DR1. Panel (a) shows the
linear-probe comparison and panel (b) shows the full fine-tuning comparison. In both panels, the selected STRADAViT
configuration is compared against the two reference baselines, ViT-MAE and DINOv2-Base(R), labeled ``DINOv2 (Registers)'' in the panels. Rows denote true
classes and columns denote predicted classes. DINOv2-Base(R) provides the stable register-based reference corresponding
to the initialization-portability experiment; it is not presented as the strongest baseline in every dataset and protocol.}\label{fig:comparison_lp_ft}
\end{figure*}

Figure~\ref{fig:comparison_lp_ft} resolves the selected model's three-fold transfer results into
classwise recall, against ViT-MAE and DINOv2-Base(R).

Under LP, RGZ DR1 differences span all six classes rather than a single category.
LoTSS DR2 is mixed: the selected model has higher FR\,I, FR\,II, and Spiral recall, while ViT-MAE
has higher Hybrid and Relaxed-double recall.

FT differences are more selective. All three models reach 70\% FR\,I recall on MiraBest.
On RGZ DR1, STRADAViT ties the highest 1C1P recall and leads 1C2P and 2C2P, whereas DINOv2-Base(R)
leads 1C3P, 2C3P, and 3C3P. These are class-specific shifts, not uniform gains.

\subsection{Latent-Space Representation Diagnostic}\label{results_umap}
\begin{figure*}[tb]
\centering
\includegraphics[width=\textwidth]{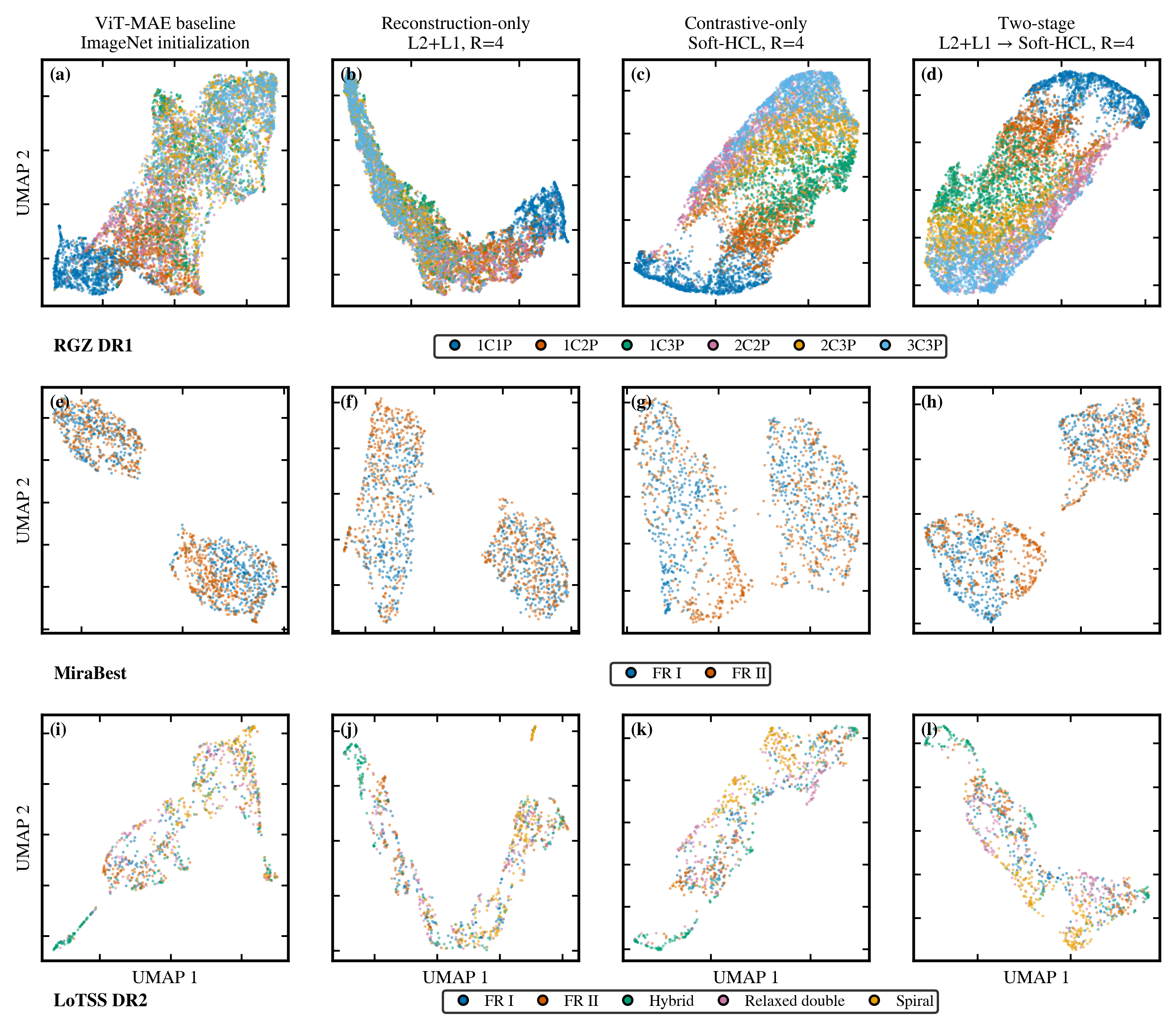}
\caption{UMAP projections of frozen embeddings for RGZ DR1 (panels (a)--(d)), MiraBest (panels (e)--(h)), and LoTSS DR2
(panels (i)--(l)). Columns follow the architecture-matched register-based pathway used for the selected release configuration:
the unadapted ViT-MAE initialization, L2+L1 reconstruction-only training, Soft-HCL contrastive-only training, and L2+L1
followed by Soft-HCL two-stage training; the adapted checkpoints use four register tokens. For each dataset, the full labeled
sample is class-balanced once with subset seed 2026 and reused across all four models; no evaluation fold is chosen based on
visual separability. The resulting samples contain 8,898 RGZ DR1, 1,490 MiraBest, and 910 LoTSS DR2 images. Each panel is
fitted independently to the unnormalized pooled embeddings using cosine distance, 30 neighbors, minimum distance 0.1, and
UMAP seed 2026. Only within-panel class separation and mixing are interpreted.}
\label{fig:representation_umap}
\end{figure*}

Figure~\ref{fig:representation_umap} compares frozen embeddings before adaptation and after the three
training branches, using matched, class-balanced samples within each dataset and UMAP \citep{McInnes2018UMAP}.

On RGZ DR1, 1C1P and 3C3P occupy more distinct regions after contrastive training, while other categories
form clearer class gradients. MiraBest shows stronger local FR\,I/FR\,II organization but substantial overlap;
LoTSS DR2 remains extensively intermixed. These independently fitted nonlinear UMAP projections are used as qualitative
companions to the transfer measurements, rather than as a model-ranking criterion.

\subsection{Overall Interpretation and Limitations}\label{results_interpretation}
Improved frozen-feature transfer does not imply uniform end-to-end gains. Contrastive branches
generally exceed reconstruction alone, but stage, loss, and register rankings remain dataset-dependent.
The repeated-seed inference applies to the complete selected configuration, not to its individual ingredients;
the ROI and branch ablations are descriptive. The paired DINOv2 comparison confirms that adaptation effects depend on
the dataset and transfer protocol, with statistically supported gains and deficits. The persistent LoTSS DR2 FT deficit
therefore motivates examining the interaction between view construction and supervised adaptation,
rather than assuming that stronger LP results guarantee better FT.

The architectural comparison is restricted to ViT encoders and the downstream experiments to image classification;
CNN and hybrid architectures, source detection, and segmentation are outside the experimental scope.

ROI-aware views reduce empty crops and raise LoTSS means relative to the no-ROI control. Their crop-scale emphasis
nevertheless differs across compact, moderately extended, and larger-scale morphologies, providing context for the
dataset-dependent transfer results.
The fixed 20-epoch downstream budget also gives very different update counts: 60 on the standard MiraBest split,
460 on LoTSS DR2, and 5140 on RGZ DR1 in the saved runs. MiraBest FT uses 12 warmup steps.

Non-overlapping tiling provides a consistent image-level sampling scheme across surveys. For extended sources,
individual tiles can contain partial emission or associated components, so the reported results characterize transfer
from the extracted cutouts rather than source-complete segmentation regions.

\section{Conclusions and Future Work}
STRADAViT is a domain-adapted ViT encoder trained with mixed-survey data, ROI-aware view generation, and
controlled continued pretraining. The selected register-based L2+L1/Soft-HCL checkpoint improves Macro-F1
over ViT-MAE across 15 paired downstream seeds in all three LP settings and in MiraBest and RGZ DR1 FT.
The LoTSS DR2 FT deficit is also significant after Holm correction, ruling out a claim of uniform downstream
improvement. The tests establish seed stability on fixed partitions for the selected configuration; stage and loss
comparisons remain descriptive results from the exploratory grid.
ROI-aware sampling improves the descriptive control means. Along the selected pathway, staged reconstruction-to-contrastive
training raises mean Macro-F1 in four of six comparisons with contrastive-only training, benefiting LoTSS DR2 and
RGZ DR1 under both protocols, but not MiraBest.

The repeated-seed DINOv2 experiment shows that the adaptation procedure can be applied to another ViT initialization,
but offers no uniform benefit: three comparisons improve, two decline, and MiraBest FT remains inconclusive after Holm
correction. Improved ViT transfer also does not imply benchmark-leading
classification: STRADAViT's MiraBest standard-split FT accuracy, $84.68\pm0.99$\%, remains below stronger
published CNN and equivariant results (Appendix~\ref{app:mirabest_standard_split}).
The selected ViT-MAE-based checkpoint is available on Hugging
Face\footnote{\url{https://huggingface.co/ISSA-ML/stradavit-base}}.
Code, configuration files, dataset-construction scripts, and evaluation utilities are maintained at
\url{https://github.com/andreademarco86/stradavit}.

Future work should examine DINO-style teacher--student training with ROI-aware views, larger or multi-scale crops
that preserve extended emission, and broader cross-survey and out-of-domain evaluation.
The frozen-feature gains motivate detection and segmentation studies; task-specific FT, stronger
initializations, and CNN--ViT hybrids can test where the present transfer limitations persist.

\appendix

\section{MiraBest Standard-Split Comparison with Prior Work}\label{app:mirabest_standard_split}
We evaluate the selected checkpoint and ViT-MAE on the standard MiraBest Confident Dataset-F split of 729 training and
104 test images. The FR\,I/FR\,II counts are $348/381$ in training and $49/55$ in testing, matching the split reported by
\citet{ScaifePorter2021EquivariantCNN,Hossain2023SemiSupervisedGCNN,Andrianomena2023RGZVAE}.
These are the MiraBest runs included in Table~\ref{tab:selected_significance}, not an additional experiment or checkpoint
selection. We retain the shared preprocessing and downstream optimization protocol of Section~\ref{transfer_eval}, with
15 seeds for each model under both linear probing and fine-tuning. Macro-F1 remains the primary internal endpoint;
accuracy and test error enable comparison with the published studies in Table~\ref{tab:mirabest_literature}.
With 729 training images and an effective batch of 256, the 20-epoch schedule gives only 60 optimizer updates.
This short common budget, the input preprocessing, and the architecture differ from task-specific literature recipes.
These protocol differences provide context for the published-score comparison. Because the standard test batch also
participated in the earlier exploratory pool, the reported uncertainty characterizes downstream-seed variation
conditional on checkpoint selection.

\begin{table*}[tb]
\caption{Published MiraBest Confident results and the selected STRADAViT checkpoint. The standard split contains
729 training and 104 test images. Architecture, preprocessing, training, and test-time procedures differ between studies;
these are published-score comparisons, not paired model comparisons. SD denotes standard deviation and SE standard error;
the uncertainty definitions are retained from each study.}
\label{tab:mirabest_literature}
\centering
\footnotesize
\setlength{\tabcolsep}{3pt}
\begin{tabular*}{\tblwidth}{@{}@{\extracolsep{\fill}}>{\raggedright\arraybackslash}p{0.35\textwidth}>{\raggedright\arraybackslash}p{0.24\textwidth}>{\raggedright\arraybackslash}p{0.34\textwidth}@{}}
\toprule
Study and method & Reported result & Split and evaluation details \\
\midrule
\citet{ScaifePorter2021EquivariantCNN}: supervised $D_{16}$-equivariant CNN & Accuracy $96.56\pm1.29$\% (SD) & Standard split; five training repeats; rotated test images \\
\citet{Bowles2021Attention}: attention-gated CNN & Accuracy 92\% & MiraBest Confident test set; published point estimate \\
\citet{Hossain2023SemiSupervisedGCNN}: BYOL E(2)-equivariant G-CNN & Accuracy $97.12\pm0.40$\% (SD) & Standard split; 15 training repeats \\
\citet{Slijepcevic2024RGZFoundation}: RGZ-pretrained BYOL, fine-tuned & Error $1.92\pm0.25$\% (SE) & Standard split; 10 training repeats \\
\citet{Andrianomena2023RGZVAE}: VDVAE + extra trees & Accuracy 82\% & Standard split; frozen latent features; $64\times64$ images \\
\midrule
STRADAViT: linear probe & Accuracy $83.85\pm2.33$\% (SD) & Standard split; 15 seeds; fixed release checkpoint \\
STRADAViT: fine-tuning & Accuracy $84.68\pm0.99$\% (SD) & Standard split; 15 seeds; same checkpoint \\
\bottomrule
\end{tabular*}
\end{table*}

\begin{table*}[tb]
\caption{Repeated linear-probe (LP) and fine-tuning (FT) comparison on the fixed MiraBest Confident $729/104$ split. Each
model/protocol combination is trained 15 times with matched random seeds. Values are mean $\pm$ standard deviation;
accuracy and error are percentages, and F1 scores are on the $[0,1]$ scale. Paired Macro-F1 differences, confidence intervals,
and significance tests are reported in Table~\ref{tab:selected_significance}.}
\label{tab:mirabest_repeated}
\centering
\footnotesize
\setlength{\tabcolsep}{2.5pt}
\begin{tabular*}{\tblwidth}{@{}@{\extracolsep{\fill}}llcccccc@{}}
\toprule
Protocol & Model & Accuracy (\%) & Error (\%) & Macro-F1 & Weighted-F1 & FR\,I F1 & FR\,II F1 \\
\midrule
LP & ViT-MAE & $67.76\pm0.62$ & $32.24\pm0.62$ & $0.660\pm0.007$ & $0.665\pm0.007$ & $0.584\pm0.011$ & $0.737\pm0.004$ \\
LP & STRADAViT & $83.85\pm2.33$ & $16.15\pm2.33$ & $0.838\pm0.023$ & $0.838\pm0.024$ & $0.835\pm0.021$ & $0.841\pm0.027$ \\
\midrule
FT & ViT-MAE & $82.50\pm1.37$ & $17.50\pm1.37$ & $0.825\pm0.014$ & $0.825\pm0.014$ & $0.822\pm0.016$ & $0.827\pm0.013$ \\
FT & STRADAViT & $84.68\pm0.99$ & $15.32\pm0.99$ & $0.847\pm0.010$ & $0.847\pm0.010$ & $0.840\pm0.010$ & $0.853\pm0.010$ \\
\bottomrule
\end{tabular*}
\end{table*}

Table~\ref{tab:mirabest_repeated} shows LP improvements in both classes, narrowing under FT.
STRADAViT's LP accuracy of $83.85\pm2.33$\% is numerically above the 82\% reported for VDVAE features with
extra trees by \citet{Andrianomena2023RGZVAE} (73\% with logistic regression). This contrasts different
representations and classifiers, not a statistically established advantage.

Conversely, STRADAViT's fine-tuned error of $15.32\pm0.99$\% remains substantially above the $1.92\pm0.25$\% reported by
\citet{Slijepcevic2024RGZFoundation}, and its accuracy remains below the attention-gated and equivariant CNN results in
Table~\ref{tab:mirabest_literature}. Matching the source split makes these performance differences informative, while the
comparison jointly reflects architecture, pretraining data, preprocessing, training duration, and model selection. The
Slijepcevic et~al. uncertainty is an SE over 10 runs, whereas ours is an SD over 15 runs; the summary statistics are therefore
compared descriptively. The paired tests establish improvements relative to ViT-MAE under our
shared protocol. The literature comparison thus supports the narrower conclusion of improved ViT transfer, not
benchmark-leading MiraBest classification.


\section*{Acknowledgements}
The authors acknowledge Xjenza Malta for funding this study under the STRADA project through the Research Excellence Programme of 2024 [REP-2024-19].
The authors thank Alessio Magro and Kristian Grixti, Institute of Space Sciences and Astronomy, University of Malta, for technical support and for maintaining the GPU computing infrastructure used for the training and evaluation runs in this study.


\printcredits

\bibliographystyle{cas-model2-names}

\bibliography{refs}

@inproceedings{Vaswani2017Attention,
 author = {Vaswani, Ashish and Shazeer, Noam and Parmar, Niki and Uszkoreit, Jakob and Jones, Llion and Gomez, Aidan N and Kaiser, \L ukasz and Polosukhin, Illia},
 booktitle = {Advances in Neural Information Processing Systems},
 editor = {I. Guyon and U. Von Luxburg and S. Bengio and H. Wallach and R. Fergus and S. Vishwanathan and R. Garnett},
 publisher = {Curran Associates, Inc.},
 title = {Attention is All you Need},
 url = {https://proceedings.neurips.cc/paper_files/paper/2017/file/3f5ee243547dee91fbd053c1c4a845aa-Paper.pdf},
 volume = {30},
 year = {2017}
}

@inproceedings{Dosovitskiy2021ViT,
  author       = {Alexey Dosovitskiy and
                  Lucas Beyer and
                  Alexander Kolesnikov and
                  Dirk Weissenborn and
                  Xiaohua Zhai and
                  Thomas Unterthiner and
                  Mostafa Dehghani and
                  Matthias Minderer and
                  Georg Heigold and
                  Sylvain Gelly and
                  Jakob Uszkoreit and
                  Neil Houlsby},
  title        = {An Image is Worth 16x16 Words: Transformers for Image Recognition
                  at Scale},
  booktitle    = {9th International Conference on Learning Representations, {ICLR} 2021,
                  Virtual Event, Austria, May 3-7, 2021},
  publisher    = {OpenReview.net},
  year         = {2021},
  url          = {https://openreview.net/forum?id=YicbFdNTTy},
  bibsource    = {dblp computer science bibliography, https://dblp.org}
}

@inproceedings{He2016ResNet,
  author = {He, Kaiming and Zhang, Xiangyu and Ren, Shaoqing and Sun, Jian},
  booktitle = {Proceedings of 2016 IEEE Conference on Computer Vision and Pattern Recognition},
  doi = {10.1109/CVPR.2016.90},
  issn = {1063-6919},
  location = {Las Vegas, NV, USA},
  month = jun,
  pages = {770--778},
  publisher = {IEEE},
  series = {CVPR '16},
  title = {{Deep Residual Learning for Image Recognition}},
  url = {http://ieeexplore.ieee.org/document/7780459},
  year = {2016}
}

@InProceedings{Radford2021CLIP,
  title = 	 {Learning Transferable Visual Models From Natural Language Supervision},
  author =       {Radford, Alec and Kim, Jong Wook and Hallacy, Chris and Ramesh, Aditya and Goh, Gabriel and Agarwal, Sandhini and Sastry, Girish and Askell, Amanda and Mishkin, Pamela and Clark, Jack and Krueger, Gretchen and Sutskever, Ilya},
  booktitle = 	 {Proceedings of the 38th International Conference on Machine Learning},
  pages = 	 {8748--8763},
  year = 	 {2021},
  editor = 	 {Meila, Marina and Zhang, Tong},
  volume = 	 {139},
  series = 	 {Proceedings of Machine Learning Research},
  month = 	 {18--24 Jul},
  publisher =    {PMLR},
  url = 	 {https://proceedings.mlr.press/v139/radford21a.html}
}

@INPROCEEDINGS {He2022MAE,
author = { He, Kaiming and Chen, Xinlei and Xie, Saining and Li, Yanghao and Dollar, Piotr and Girshick, Ross },
booktitle = { 2022 IEEE/CVF Conference on Computer Vision and Pattern Recognition (CVPR) },
title = {{ Masked Autoencoders Are Scalable Vision Learners }},
year = {2022},
volume = {},
ISSN = {},
pages = {15979-15988},
doi = {10.1109/CVPR52688.2022.01553},
url = {https://doi.ieeecomputersociety.org/10.1109/CVPR52688.2022.01553},
publisher = {IEEE Computer Society},
address = {Los Alamitos, CA, USA},
month = {June}
}

@InProceedings{Chen2020SimCLR,
  title = 	 {A Simple Framework for Contrastive Learning of Visual Representations},
  author =       {Chen, Ting and Kornblith, Simon and Norouzi, Mohammad and Hinton, Geoffrey},
  booktitle = 	 {Proceedings of the 37th International Conference on Machine Learning},
  pages = 	 {1597--1607},
  year = 	 {2020},
  editor = 	 {III, Hal Daumé and Singh, Aarti},
  volume = 	 {119},
  series = 	 {Proceedings of Machine Learning Research},
  month = 	 {13--18 Jul},
  publisher =    {PMLR},
  url = 	 {https://proceedings.mlr.press/v119/chen20j.html}
}

@INPROCEEDINGS{He2020MoCo,
  author={He, Kaiming and Fan, Haoqi and Wu, Yuxin and Xie, Saining and Girshick, Ross},
  booktitle={2020 IEEE/CVF Conference on Computer Vision and Pattern Recognition (CVPR)}, 
  title={Momentum Contrast for Unsupervised Visual Representation Learning}, 
  year={2020},
  volume={},
  number={},
  pages={9726-9735},
  doi={10.1109/CVPR42600.2020.00975}
  }

@inproceedings{Grill2020BYOL,
 author = {Grill, Jean-Bastien and Strub, Florian and Altch\'{e}, Florent and Tallec, Corentin and Richemond, Pierre and Buchatskaya, Elena and Doersch, Carl and Avila Pires, Bernardo and Guo, Zhaohan and Gheshlaghi Azar, Mohammad and Piot, Bilal and kavukcuoglu, koray and Munos, Remi and Valko, Michal},
 booktitle = {Advances in Neural Information Processing Systems},
 editor = {H. Larochelle and M. Ranzato and R. Hadsell and M.F. Balcan and H. Lin},
 pages = {21271--21284},
 publisher = {Curran Associates, Inc.},
 title = {Bootstrap Your Own Latent - A New Approach to Self-Supervised Learning},
 url = {https://proceedings.neurips.cc/paper_files/paper/2020/file/f3ada80d5c4ee70142b17b8192b2958e-Paper.pdf},
 volume = {33},
 year = {2020}
}

@inproceedings{Bardes2022VICReg,
  TITLE = {{VICReg: Variance-Invariance-Covariance Regularization For Self-Supervised Learning}},
  AUTHOR = {Bardes, Adrien and Ponce, Jean and Lecun, Yann},
  URL = {https://inria.hal.science/hal-03541297},
  BOOKTITLE = {{ICLR 2022 - International Conference on Learning Representations}},
  ADDRESS = {Online, United States},
  YEAR = {2022},
  MONTH = {April},
  HAL_ID = {hal-03541297},
  HAL_VERSION = {v1},
}

@article{Caron2021DINO,
  title={Emerging Properties in Self-Supervised Vision Transformers},
  author={Mathilde Caron and Hugo Touvron and Ishan Misra and Herv{\'e} J{\'e}gou and Julien Mairal and Piotr Bojanowski and Armand Joulin},
  journal={2021 IEEE/CVF International Conference on Computer Vision (ICCV)},
  year={2021},
  pages={9630-9640},
  url={https://api.semanticscholar.org/CorpusID:233444273}
}

@misc{Oquab2023DINOv2,
  author       = {Oquab, Maxime and Darcet, Timoth{\'e}e and Moutakanni, Th{\'e}o and Vo, Huy and Szafraniec, Marc and Khalidov, Vasil and Fernandez, Pierre and Haziza, Daniel and Massa, Francisco and El-Nouby, Alaaeldin and Assran, Mahmoud and Ballas, Nicolas and Galuba, Wojciech and Howes, Russell and Huang, Po-Yao and Li, Shang-Wen and Misra, Ishan and Rabbat, Michael and Sharma, Vasu and Synnaeve, Gabriel and Xu, Hu and Jegou, Herv{\'e} and Mairal, Julien and Labatut, Patrick and Joulin, Armand and Bojanowski, Piotr},
  title        = {DINOv2: Learning Robust Visual Features without Supervision},
  howpublished = {arXiv:2304.07193},
  year         = {2023},
  archivePrefix= {arXiv},
  eprint       = {2304.07193},
  doi          = {10.48550/arXiv.2304.07193},
  url          = {https://arxiv.org/abs/2304.07193}
}

@ARTICLE{Horton2025LoTSSDR2Morphology,
       author = {{Horton}, M.~A. and {Hardcastle}, M.~J. and {Miley}, G.~K. and {Tasse}, C. and {Shimwell}, T.~W.},
        title = {Complex morphology and precession indicators of active galactic nuclei jets in LoTSS DR2},
      journal = {Astronomy and Astrophysics},
         year = {2025},
        month = {March},
       volume = {699},
          eid = {A338},
        pages = {A338},
          doi = {10.1051/0004-6361/202453559},
       adsurl = {https://ui.adsabs.harvard.edu/abs/2025A&A...699A.338H}
}

@article{PorterScaife2023MiraBest,
    author = {Porter, Fiona A M and Scaife, Anna M M},
    title = {MiraBest: a data set of morphologically classified radio galaxies for machine learning},
    journal = {RAS Techniques and Instruments},
    volume = {2},
    number = {1},
    pages = {293-306},
    year = {2023},
    month = {06},
    issn = {2752-8200},
    doi = {10.1093/rasti/rzad017},
    url = {https://doi.org/10.1093/rasti/rzad017},
}

@article{Slijepcevic2024RGZFoundation,
    author = {Slijepcevic, Inigo V and Scaife, Anna M M and Walmsley, Mike and Bowles, Micah and Wong, O Ivy and Shabala, Stanislav S and White, Sarah V},
    title = {Radio galaxy zoo: towards building the first multipurpose foundation model for radio astronomy with self-supervised learning},
    journal = {RAS Techniques and Instruments},
    volume = {3},
    number = {1},
    pages = {19-32},
    year = {2023},
    month = {12},
    issn = {2752-8200},
    doi = {10.1093/rasti/rzad055},
    url = {https://doi.org/10.1093/rasti/rzad055},
}

@article{RGZDR1Zenodo14195049,
    author = {Wong, O Ivy and Garon, A F and Alger, M J and Rudnick, L and Shabala, S S and Willett, K W and Banfield, J K and Andernach, H and Norris, R P and Swan, J and Hardcastle, M J and Lintott, C J and White, S V and Seymour, N and Kapińska, A D and Tang, H and Simmons, B D and Schawinski, K},
    title = {Radio Galaxy Zoo data release 1: 100185 radio source classifications from the FIRST and ATLAS surveys},
    journal = {Monthly Notices of the Royal Astronomical Society},
    volume = {536},
    number = {4},
    pages = {3488-3506},
    year = {2024},
    month = {12},
    issn = {0035-8711},
    doi = {10.1093/mnras/stae2790},
    url = {https://doi.org/10.1093/mnras/stae2790},
}

@article{Riggi2016CAESAR,
    author = {Riggi, S. and Ingallinera, A. and Leto, P. and Cavallaro, F. and Bufano, F. and Schillirò, F. and Trigilio, C. and Umana, G. and Buemi, C. S. and Norris, R. P.},
    title = {Automated detection of extended sources in radio maps: progress from the SCORPIO survey},
    journal = {Monthly Notices of the Royal Astronomical Society},
    volume = {460},
    number = {2},
    pages = {1486-1499},
    year = {2016},
    month = {04},
    issn = {0035-8711},
    doi = {10.1093/mnras/stw982},
    url = {https://doi.org/10.1093/mnras/stw982},
}

@article{Riggi2019CaesarPASA,
  title   = {The {CAESAR} source finder: recent developments and testing},
  author={Simone Riggi and Fabio Vitello and Ugo Becciani and Carla S. Buemi and Filomena Bufano and Antonio Calanducci and Francesco Cavallaro and Allison H. Costa and Adriano Ingallinera and Paolo Leto and Sara Loru and Ray P. Norris and F Schillir{\`o} and Eva Sciacca and Corrado Trigilio and Grazia Umana},
  journal={Publications of the Astronomical Society of Australia},
  year={2019},
  volume={36},
  pages={e037},
  doi={10.1017/pasa.2019.29},
  url={https://doi.org/10.1017/pasa.2019.29}
}

@article{Riggi2024PASAContrastive,
       author = {Riggi, S. and Cecconello, T. and Palazzo, S. and Hopkins, A.~M. and Gupta, N. and Bordiu, C. and Ingallinera, A. and Buemi, C. and Bufano, F. and Cavallaro, F. and Filipovi{\'c}, M.~D. and Leto, P. and Loru, S. and Ruggeri, A.~C. and Trigilio, C. and Umana, G. and Vitello, F.},
        title = "{Self-supervised contrastive learning of radio data for source detection, classification and peculiar object discovery}",
      journal = {PASA},
         year = {2024},
        month = {November},
       volume = {41},
          eid = {e085},
        pages = {e085},
          doi = {10.1017/pasa.2024.84},
archivePrefix = {arXiv},
       eprint = {2404.18462},
 primaryClass = {astro-ph.IM}
}

@misc{Riggi2024DeepLearningRadio,
  author       = {Riggi, Simone and Cecconello, Thomas and Hopkins, Andrew M. and Trigilio, Corrado and Umana, Grazia},
  title        = {Detection and classification of radio sources with deep learning},
  howpublished = {arXiv:2411.08519},
  year         = {2024},
  doi          = {10.48550/arXiv.2411.08519},
  url          = {https://arxiv.org/abs/2411.08519}
}

@InProceedings{Cecconello2025RadioSSLBenchmark,
  author    = {Cecconello, Thomas and Riggi, Simone and Becciani, Ugo and Vitello, Francesco and Hopkins, Andrew M. and Vizzari, Giuseppe and Spampinato, Concetto and Palazzo, Simone},
  title     = {Self-supervised Learning for Radio Astronomy Source Classification: a Benchmark},
  booktitle= {Pattern Recognition. ICPR 2024 International Workshops and Challenges},
  year={2025},
  series    = {Lecture Notes in Computer Science},
  publisher = {Springer},
  pages={424--439},
  isbn={978-3-031-88217-3}
}

@article{Riggi2025RadioLLaVA,
  author  = {Riggi, Simone and Cecconello, Thomas and Pilzer, Andrea and Palazzo, Simone and Gupta, Nikhel and Hopkins, Andrew M. and Trigilio, Corrado and Umana, Grazia},
  title   = {{Radio-LLaVA}: advancing vision-language models for radio astronomical source analysis},
  journal = {Publications of the Astronomical Society of Australia},
  year    = {2025},
  volume  = {42},
  pages   = {121},
  doi     = {10.1017/pasa.2025.10082},
}

@article{BaronPerez2025AandASSL,
  author  = {Baron P{\'e}rez, Nicolas and Br{\"u}ggen, Marcus and Kasieczka, Gregor and Lucie-Smith, Luisa},
  title   = {Classification of radio sources through self-supervised learning},
  journal = {Astronomy \& Astrophysics},
  year    = {2025},
  volume  = {699},
  pages   = {A302},
  doi     = {10.1051/0004-6361/202554735},
  url     = {https://www.aanda.org/articles/aa/full_html/2025/07/aa54735-25/aa54735-25.html}
}

@article{Lastufka2024MeerKATSSL,
  author  = {Lastufka, Erica and Bait, Omkar and Taran, Olga and Drozdova, Mariia and Kinakh, Vitaliy and Piras, Davide and Audard, Marc and Dessauges-Zavadsky, Miroslava and Holotyak, Taras and Schaerer, Daniel and Voloshynovskiy, Slava},
  title   = {Self-supervised learning on {MeerKAT} wide-field continuum images},
  journal = {Astronomy \& Astrophysics},
  year    = {2024},
  volume  = {690},
  pages   = {A310},
  doi     = {10.1051/0004-6361/202449964},
  url     = {https://www.aanda.org/articles/aa/full_html/2024/10/aa49964-24/aa49964-24.html}
}

@article{Buatthaisong2025MNRASSelfSupervisedFR,
  author  = {Buatthaisong, Nutthawara and Slijepcevic, Inigo Val and Scaife, Anna M. M. and Bowles, Micah and Hopkins, Andrew M. and Mohan, Devina and Shabala, Stanislav S. and Wong, O. Ivy},
  title   = {{Radio Galaxy Zoo}: morphological classification by Fanaroff--Riley designation using self-supervised pre-training},
  journal = {Monthly Notices of the Royal Astronomical Society},
  year    = {2025},
  volume  = {544},
  number  = {4},
  pages   = {4062--4078},
  doi     = {10.1093/mnras/staf1942},
  url     = {https://academic.oup.com/mnras/article-abstract/544/4/4062/8139979}
}

@misc{Drozdova2025VLMRadio,
  author       = {Drozdova, Mariia and Lastufka, Erica and Kinakh, Vitaliy and Holotyak, Taras and Schaerer, Daniel and Voloshynovskiy, Slava},
  title        = {Radio Astronomy in the Era of Vision--Language Models: Prompt Sensitivity and Adaptation},
  howpublished = {arXiv:2509.02615},
  year         = {2025},
  doi          = {10.48550/arXiv.2509.02615},
  url          = {https://arxiv.org/abs/2509.02615}
}

@article{Bonaldi2020,
    author = {Bonaldi, A and An, T and Brüggen, M and Burkutean, S and Coelho, B and Goodarzi, H and Hartley, P and Sandhu, P K and Wu, C and Yu, L and Zhoolideh Haghighi, M H and Antón, S and Bagheri, Z and Barbosa, D and Barraca, J P and Bartashevich, D and Bergano, M and Bonato, M and Brand, J and de Gasperin, F and Giannetti, A and Dodson, R and Jain, P and Jaiswal, S and Lao, B and Liu, B and Liuzzo, E and Lu, Y and Lukic, V and Maia, D and Marchili, N and Massardi, M and Mohan, P and Morgado, J B and Panwar, M and Prabhakar, P and Ribeiro, V A R M and Rygl, K L J and Sabz Ali, V and Saremi, E and Schisano, E and Sheikhnezami, S and Vafaei Sadr, A and Wong, A and Wong, O I},
    title = {Square Kilometre Array Science Data Challenge 1: analysis and results},
    journal = {Monthly Notices of the Royal Astronomical Society},
    volume = {500},
    number = {3},
    pages = {3821-3837},
    year = {2020},
    month = {10},
    issn = {0035-8711},
    doi = {10.1093/mnras/staa3023},
    url = {https://doi.org/10.1093/mnras/staa3023},
}

@inproceedings{Robinson2021HardNegatives,
  author       = {Joshua David Robinson and
                  Ching{-}Yao Chuang and
                  Suvrit Sra and
                  Stefanie Jegelka},
  title        = {Contrastive Learning with Hard Negative Samples},
  booktitle    = {9th International Conference on Learning Representations, {ICLR} 2021,
                  Virtual Event, Austria, May 3-7, 2021},
  publisher    = {OpenReview.net},
  year         = {2021},
  url          = {https://openreview.net/forum?id=CR1XOQ0UTh-},
  bibsource    = {dblp computer science bibliography, https://dblp.org}
}

@article{Andrianomena2023RGZVAE,
  author        = {Andrianomena, Sambatra and Tang, Hongming},
  title         = {Radio Galaxy Zoo: Leveraging latent space representations from variational autoencoder},
  journal       = {arXiv preprint arXiv:2311.08331},
  year          = {2023},
  eprint        = {2311.08331},
  archivePrefix = {arXiv},
  primaryClass  = {astro-ph.IM},
  url           = {https://arxiv.org/abs/2311.08331}
}

@article{Hossain2023SemiSupervisedGCNN,
  author       = {Hossain, Mir Sazzat and Roy, Sugandha and Asad, K. M. B. and Momen, Arshad and Ali, Amin Ahsan and Amin, M. Ashraful and Rahman, A. K. M. Mahbubur},
  title        = {Morphological Classification of Radio Galaxies using Semi-Supervised Group Equivariant {CNN}s},
  journal      = {Procedia Computer Science},
  volume       = {222},
  pages        = {601--612},
  year         = {2023},
  doi          = {10.1016/j.procs.2023.08.198},
  url          = {https://doi.org/10.1016/j.procs.2023.08.198}
}

@article{Bowles2021Attention,
  author       = {Bowles, Micah and Scaife, Anna M. M. and Porter, Fiona and Tang, Hongming and Bastien, David J.},
  title        = {Attention-gating for Improved Radio Galaxy Classification},
  journal      = {Monthly Notices of the Royal Astronomical Society},
  volume       = {501},
  number       = {3},
  pages        = {4579--4595},
  year         = {2021},
  doi          = {10.1093/mnras/staa3946},
  url          = {https://doi.org/10.1093/mnras/staa3946}
}

@article{ScaifePorter2021EquivariantCNN,
  author       = {Scaife, Anna M. M. and Porter, Fiona},
  title        = {Fanaroff--Riley Classification of Radio Galaxies using Group-Equivariant Convolutional Neural Networks},
  journal      = {Monthly Notices of the Royal Astronomical Society},
  volume       = {503},
  number       = {2},
  pages        = {2369--2379},
  year         = {2021},
  doi          = {10.1093/mnras/stab530},
  url          = {https://doi.org/10.1093/mnras/stab530}
}

@article{Knowles2022MGCLS,
  author = {Knowles, K. and Cotton, W. D. and Rudnick, L. and others},
  title = {The {MeerKAT} Galaxy Cluster Legacy Survey. I. Survey overview and highlights},
  journal = {Astronomy \& Astrophysics},
  year = {2022},
  volume = {657},
  pages = {A56},
  doi = {10.1051/0004-6361/202141488}
}

@article{Shimwell2022LoTSS,
  author = {Shimwell, T. W. and Hardcastle, M. J. and Tasse, C. and others},
  title = {The {LOFAR} Two-metre Sky Survey. V. Second data release},
  journal = {Astronomy \& Astrophysics},
  year = {2022},
  volume = {659},
  pages = {A1},
  url = {https://arxiv.org/abs/2202.11733}
}

@article{Becker1995FIRST,
  author = {Becker, Robert H. and White, Richard L. and Helfand, David J.},
  title = {The {FIRST} Survey: Faint Images of the Radio Sky at Twenty Centimeters},
  journal = {The Astrophysical Journal},
  year = {1995},
  volume = {450},
  pages = {559},
  doi = {10.1086/176166}
}

@article{Astropy2013,
  author = {{Astropy Collaboration} and Robitaille, T. P. and Tollerud, E. J. and others},
  title = {Astropy: A community {Python} package for astronomy},
  journal = {Astronomy \& Astrophysics},
  year = {2013},
  volume = {558},
  pages = {A33},
  doi = {10.1051/0004-6361/201322068}
}

@article{Planck2020Parameters,
  author = {{Planck Collaboration} and Aghanim, N. and Akrami, Y. and others},
  title = {{Planck} 2018 results. VI. Cosmological parameters},
  journal = {Astronomy \& Astrophysics},
  year = {2020},
  volume = {641},
  pages = {A6},
  doi = {10.1051/0004-6361/201833910}
}

@inproceedings{Darcet2024Registers,
  author = {Darcet, Timoth{\'e}e and Oquab, Maxime and Mairal, Julien and Bojanowski, Piotr},
  title = {Vision Transformers Need Registers},
  booktitle = {International Conference on Learning Representations},
  year = {2024},
  url = {https://openreview.net/forum?id=2dnO3LLiJ1}
}

@inproceedings{Loshchilov2019AdamW,
  author = {Loshchilov, Ilya and Hutter, Frank},
  title = {Decoupled Weight Decay Regularization},
  booktitle = {International Conference on Learning Representations},
  year = {2019},
  url = {https://arxiv.org/abs/1711.05101}
}

@misc{McInnes2018UMAP,
  author = {McInnes, Leland and Healy, John and Melville, James},
  title = {{UMAP}: Uniform Manifold Approximation and Projection for Dimension Reduction},
  year = {2018},
  howpublished = {arXiv:1802.03426},
  url = {https://arxiv.org/abs/1802.03426}
}

@article{Holm1979,
  author = {Holm, Sture},
  title = {A Simple Sequentially Rejective Multiple Test Procedure},
  journal = {Scandinavian Journal of Statistics},
  year = {1979},
  volume = {6},
  number = {2},
  pages = {65--70},
  url = {https://www.jstor.org/stable/4615733}
}



\end{document}